\documentclass[11pt]{myArticle}       
\pdfoutput=1

\usepackage{amsmath,amssymb,amsfonts} 
\usepackage{graphicx}
\usepackage{epsfig}
\usepackage{color}                    
\usepackage{hyperref}                 
\usepackage{myFancyhdr}               
\usepackage{makeidx}                  
\usepackage{helvet}                   
\usepackage{setspace}                 
\usepackage{fancybox}                 
\usepackage{float}                    
\usepackage{xspace}                   
\usepackage{wasysym}                  
\usepackage[numbers,sort&compress]{natbib}
\usepackage{hypernat}
\usepackage{wrapfig}
\usepackage{subcaption}
\usepackage{lineno}
\usepackage{listings}
\usepackage{multirow}

\definecolor{mygray}{rgb}{0.3,0.32,0.35}
\definecolor{darkblue1}{rgb}{0,0,.2}
\definecolor{darkblue}{rgb}{0,0,.3}
\definecolor{darkred}{rgb}{0.5,0,0}
\pagecolor{white} 
\color{black}     
\hypersetup{breaklinks=true,
            colorlinks=true,
            linkcolor=blue,
            menucolor=blue,
            urlcolor=blue,
            citecolor=blue,
}

\newcommand\met{\ensuremath{E_\mathrm{T}^{\mathrm{miss}}}}
\newcommand{\ttbar} {\ensuremath{t\overline t}\xspace}
\bibstyle{plain}
\newcommand\allFontSize{\small}

\newenvironment{myquote}
               {\list{}{\leftmargin0cm}%
                \item\relax}
               {\endlist}
\usepackage{caption}

\newcommand\detailsSize{\allFontSize}
\newenvironment{details}%
{\begin{myquote}\vspace{-0.2cm}\detailsSize}{\end{myquote}\vspace{-0.2cm}}

\makeindex
\begin{document}
%
%

\pagenumbering{arabic}
{\small
\color{mygray}
\begin{flushright}
{\sf\em \today} \\
\url{http://cern.ch/histfitter}
\end{flushright}
}
\def\UrlFont{\rm}

\vspace{1.3cm}

{\sf\LARGE\bfseries
HistFitter software framework for statistical data analysis}

\vspace{1.0cm}

{\Large
  M.~Baak$^{a}$, G.J.~Besjes$^{b,c}$, D.~C\^ot\'e$^{d}$, A.~Koutsman$^{e}$, J.~Lorenz$^{f,g}$ and D.~Short$^{h}$
}

\vspace{0.5cm}

{\normalsize
  $^{a}$ CERN, Geneva, Switzerland \\
  ${^b}$ Experimental and Theoretical High Energy Physics, IMAPP, Faculty of Science, Radboud University Nijmegen, The Netherlands \\
  ${^c}$ Nikhef, Amsterdam, The Netherlands \\
  $^{d}$ University of Texas, Arlington, USA \\
  $^{e}$ TRIUMF, Vancouver, Canada \\
  $^{f}$ Ludwig-Maximilians-Universit\"at M\"unchen, M\"unchen, Germany \\
  $^{g}$ Excellence Cluster Universe, Garching, Germany \\
  $^{h}$ University of Oxford, Oxford, UK

\vspace{1.0cm}

\begin{details} {\sf\bfseries Abstract}

We present a software framework for statistical data analysis, called HistFitter, 
that has been used extensively by the ATLAS Collaboration to analyze big datasets 
originating from proton-proton collisions at the Large Hadron Collider at CERN. 
Since 2012 HistFitter has been the standard statistical tool in searches for supersymmetric
particles performed by ATLAS.

HistFitter is a programmable and flexible framework to build, book-keep, fit, interpret and present
results of data models of nearly arbitrary complexity.
Starting from an object-oriented configuration, defined by users, the framework  
builds probability density functions that are automatically
fitted to data and interpreted with statistical tests.
A key innovation of HistFitter is its design, which is rooted in core analysis strategies of particle physics.
The concepts of control, signal and validation regions are woven into its very fabric.
These are progressively treated with statistically rigorous built-in methods.
Being capable of working with multiple data models at once, HistFitter introduces 
an additional level of abstraction that allows for easy bookkeeping,
manipulation and testing of large collections of signal hypotheses.
Finally, HistFitter provides a collection of tools to present results 
with publication-quality style through a simple command-line interface.

\end{details}


\thispagestyle{empty}

\newpage
%
%
\tableofcontents
\newpage

\section{Introduction}
\label{sec:intro}

This paper describes a software framework for statistical data analysis, called ``HistFitter'', 
that has been used extensively by the ATLAS Collaboration~\cite{ATLAS} to analyze
big datasets originating from proton-proton collisions at the Large Hadron Collider (LHC) at CERN. 
Most notably, HistFitter has become a \emph{de facto} standard in searches for supersymmetric particles since 2012, 
see for example~\cite{PhysRevD.86.092002,PhysRevD.87.012008,Aad:2012ywa,Aad:2012xqa,Aad:2012pxa,Aad:2012hba,Aad:2012yr,Aad:2012uu,ATLAS:2012ht,ATLAS:2012kr,Aad:2012naa,Aad:2013wta,Aad:2013ija,Aad:2014nua,Aad:2014qaa,Aad:2014mha,Aad:2014vma,Aad:2014pda}, with 
some usage for Exotic~\cite{ATLAS:2012ky,Aad:2014gka} and Higgs boson~\cite{HbbCONF} physics. HistFitter is written in Python and C++, 
the former being used for configuration and the latter for CPU-intensive calculations. 
Internally, HistFitter uses the software packages 
HistFactory~\cite{Cranmer:2012sba} and RooStats~\cite{Moneta:2010pm},
which are based on RooFit~\cite{Verkerke:2003ir} and ROOT~\cite{Brun:1997pa,Antcheva:2011zz},
to construct parametric models and perform statistical tests of the data.
HistFitter extends these tools in four key areas:
\begin{enumerate}
\item \textbf{Programmable framework:} HistFitter performs complete statistical analyses 
of pre-formatted input data samples, from a single user-defined configuration file,
by putting together tools from several sources in a coherent and programmable framework.
\item \textbf{Analysis strategy:} HistFitter has built-in concepts of control, signal and 
validation regions, which are used to constrain, extrapolate and validate data model 
predictions across an analysis. The framework also introduces a statistically rigorous treatment of the validation regions. 
\item \textbf{Bookkeeping:} HistFitter can keep track of numerous data models, 
including all generated input histograms, both before and after adjustment to measured data, 
and can perform statistical tests and model-parameter scans of all these models
in an organized way. This introduces a powerful additional level of abstraction, 
which aids the processing of large collections of signal hypothesis tests.
\item \textbf{Presentation and interpretation:} HistFitter provides a collection of methods to 
determine the statistical significance of signal hypotheses,
estimate the quality of likelihood fits,
and produce publication-quality tables and plots expressing these results. 
\end{enumerate}

This paper details these extensions and is organized as follows. 
Section~\ref{sec:analysisStrategy} summarizes the data analysis strategy and the statistical formalism used
for many searches and measurements at the LHC. 
Section~\ref{sec:hfframework} describes how this strategy is ingrained in the HistFitter framework.
Section~\ref{sec:pdf} sketches how support for multiple Probability Density Function (PDF) instances of 
nearly arbitrary complexity
has been implemented in HistFitter with a modular object-oriented design. 
Section~\ref{sec:fit} explains how the PDFs can be used to perform statistical fits\footnote{Also referred to as 
``regressions'' in the literature.} of various types.
Section~\ref{sec:pres} describes how the fit results can be conveniently presented and visualized with 
different methods.
Finally, Sec.~\ref{sec:interpret} shows how signal hypotheses can be tested quantitatively in several ways.
The publicly available release of HistFitter is described in Sec.~\ref{sec:implementation}, 
before concluding in Sec.~\ref{sec:conclu}.

\section{Data analysis strategy}
\label{sec:analysisStrategy}

Particle physics experiments require the careful analysis of large data samples, 
coming from an experimental apparatus, in order to measure the properties of fundamental particles.
A very active field of research is focused on using these datasets to discover physical processes 
that have been predicted by theoretical models, but have not yet been observed in nature.

Analyses generally rely on external predictions for the various background and signal components 
in the data to aid the interpretation of observations,
where the signal component describes the process of interest.
In particle physics, simulations of known and hypothesized physics processes are run through a 
detailed detector simulation, and are subsequently reconstructed with the same algorithms as the data.
In addition, background samples can be constructed using data-driven methods.
The simulated samples may depend on one or many model parameters, 
for example the masses of hypothesized new particles
such as foreseen by supersymmetry.
It may be required, for instance when signals are analyzed over a multi-dimensional space of model parameters, 
to sample from a ``grid'' of potential signal scenarios, 
with each point on that grid corresponding to a unique point in the multi-dimensional parameter space.
If no excess is observed in the data, exclusion limits may be set within this grid, excluding a subset of the tested parameter values. 

HistFitter configures and builds parametric models to describe the observed data, 
and provides tools to interpret the data in terms of these models. 
It uses the concepts of control, validation, and signal regions in the construction and handling of these models.
A key innovation of HistFitter is to weave these concepts into its very fabric, 
and to treat them with statistically rigorous methods.
The technical implementation of HistFitter is detailed in the following sections, 
where we explain two key ideas in data analysis strategy that have helped shape HistFitter.

\subsection{Use of control, signal and validation regions}

Any physics analysis aiming to study a specific phenomenon involves defining a region of phase space,
obtained by applying selections to a set of kinematic observables,
where a particular signal model predicts a significant excess of events over the predicted background level. 
Such a signal enriched region is called a \emph{signal region}, or SR.

To estimate background processes contaminating the SR(s) in a semi-data-driven way, one typically defines 
\emph{control region(s)}, or CR(s), in which the dominant background(s)
can be controlled by comparison to the data samples.
CRs are specifically designed to have a high purity for one type of background,
and should be free of signal contamination.

A third important component of data analysis 
is the validation of the model used to predict the number of background events in the SR(s). 
\emph{Validation region(s)}, VR(s), are defined for this purpose.
VR(s) are typically placed in between the CR(s) and SR(s).
Hence, the choice of VR(s) is typically a trade-off between maximizing statistical significance and 
minimizing signal contamination, while controlling the assumptions in the extrapolation
from CR(s) to SR(s).

\begin{figure}[h!]
\centering
\includegraphics[width=0.65\textwidth]{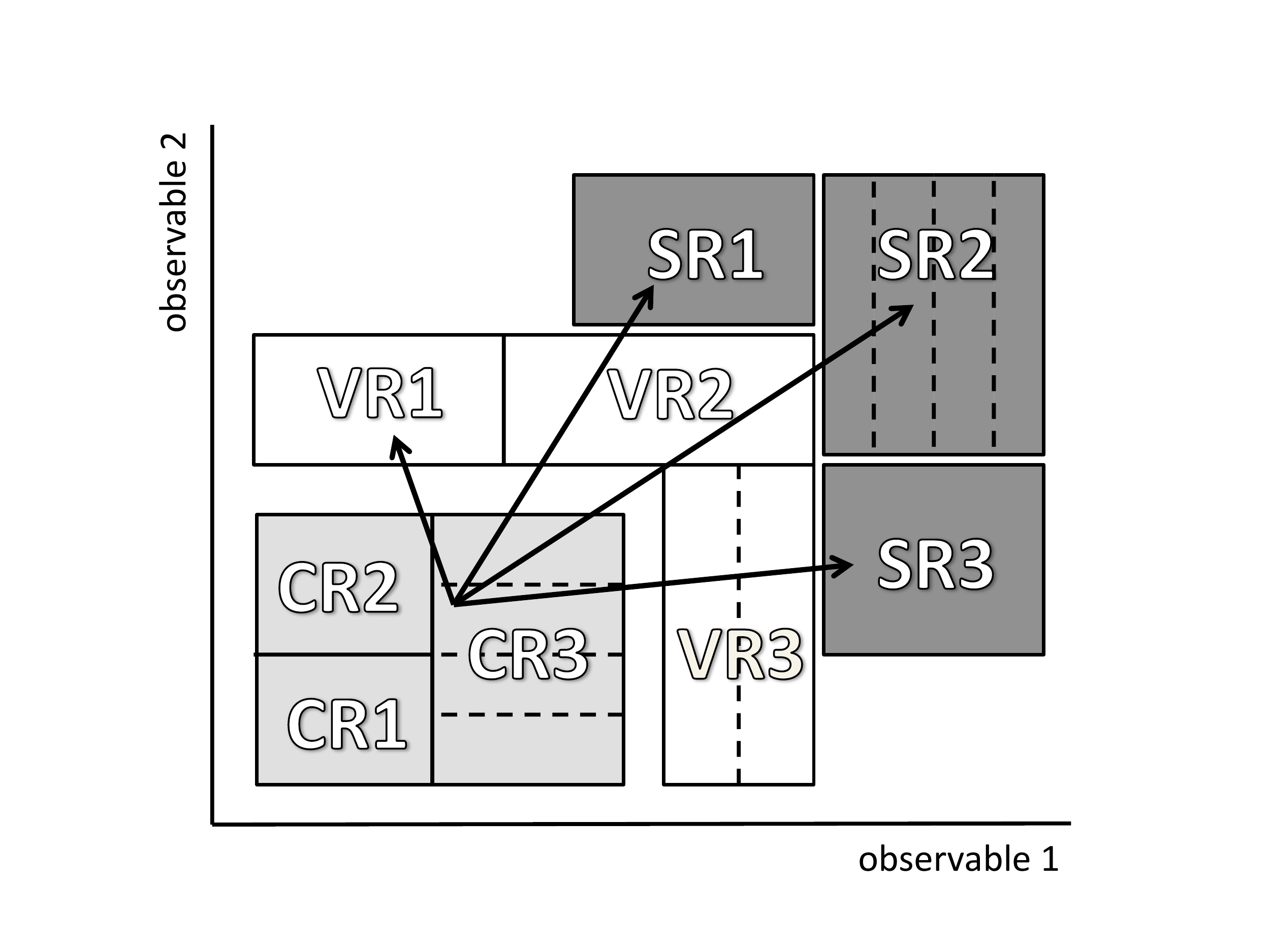}
\caption{A schematic view of an analysis strategy with multiple control, validation and signal regions. 
All regions can have single- or multiple bins, as illustrated by the dashed lines. 
The extrapolation from the control to the signal regions is verified in the validation regions 
that lie in the extrapolation phase space.}
\label{fig:cartoon_CRVRSR}
\end{figure}
%
The concept of extrapolation between CRs, VRs and SRs is 
schematically shown in Fig.~\ref{fig:cartoon_CRVRSR}.
Any such region can have one or many event bins, as illustrated by the dashed lines. 
The extrapolation happens in observables chosen to separate the regions, as discussed 
in Sec.~\ref{sec:intro_extrapolation}, and shown by the arrows on the figure.

To extract accurate and quantitative information from the data,
particle physicists frequently use a Probability Density Function (PDF) 
whose parameters are adjusted with a fitting procedure.
The fit to data is based on statistically independent CRs and SRs,
which ensures that they can be modeled by separate PDFs
and combined into a simultaneous fit.
A crucial point of the HistFitter analysis strategy is the sharing of PDF parameters 
in all regions: CRs, SRs and VRs.
This procedure enables the use of information from each signal and background component, as well as
systematics uncertainties, consistently in all regions.

The analysis strategy flow is schematically shown in Fig.~\ref{fig:overview}.
Through the fit to data, the observed event counts in CR(s) are used to coherently 
normalize background estimates in all regions, notably the SR(s).
If the dominant background processes are estimated with Monte Carlo (MC) simulations,
their initial predictions are scaled to observed levels in the corresponding CRs 
using \emph{normalization factors} computed in the fit. 
This results in so-called ``normalized background predictions''. These are then used for extrapolation
into the VRs and SRs,
as discussed in the next sub-section.

%
\begin{figure}[tbh!]
\centering
\includegraphics[width=0.95\textwidth]{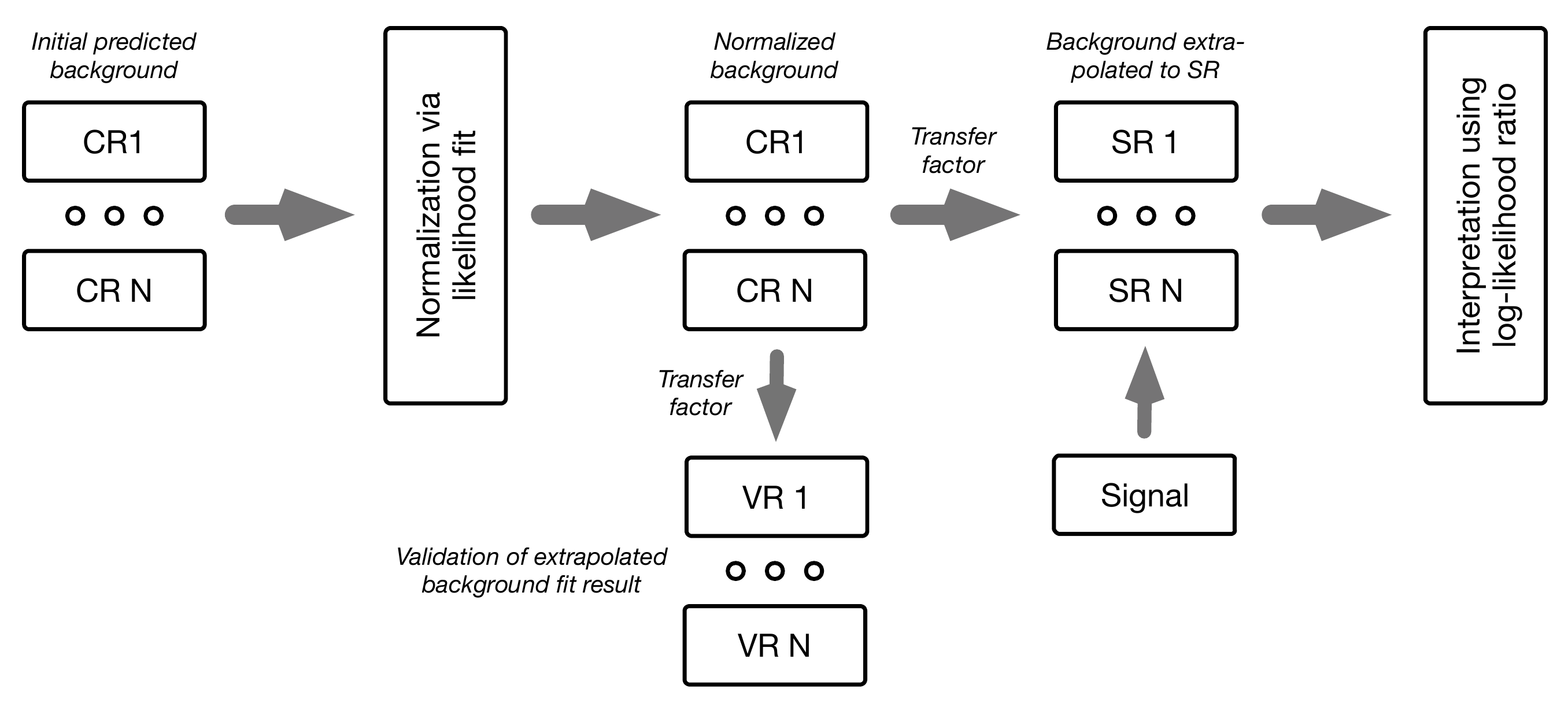}
\caption{Overview of a typical analysis strategy flow with HistFitter. }
\label{fig:overview}
\end{figure}

\subsection{Extrapolation and transfer factors}
\label{sec:intro_extrapolation}

An underlying assumption has been made in the previous sections, notably
that extrapolations over the kinematic variables 
used to differentiate SR(s) from CR(s) are well modeled after fitting the PDF to data in CR(s). 
Once the dominant background processes have been normalized in CR(s),
the corresponding modifications to the PDF can be extrapolated to the VR(s),
which is (are) then used to verify the validity of this assumption.
In HistFitter, this extrapolation is statistically rigorous because the PDF is coherently defined in 
all the CRs, SRs and VRs, even though the VRs are not used as a constraint in the fit. 
Technical details on the comprehensive extrapolation technique used in HistFitter
are given in Sec.~\ref{sec:errorpropagation}.

Once a satisfactory agreement is found between normalized background predictions 
and observed data in the VRs, the background predictions are further extrapolated to the SR(s),
and, by convention, are only then compared with the observed data (see Fig.~\ref{fig:overview});
a process generally called ``unblinding'' or ``opening the box''.
This unblinding strategy is useful for validating the performance of the extrapolations,
and, in a wider physics context, a) to get confidence in the methods used
and b) to avoid analyzers from using premature SR predictions, and thus potentially biasing the physics results.

Key ingredients of the fitting procedure are the ratios of expected event counts, 
called \emph{transfer factors}, or TFs, 
of each normalized background process between each SR and each CR.
The TFs allow the observations in the CRs to be converted into background estimates in the SRs, using:
{\small
\begin{equation}
\label{eq:tf}
N_p({\rm SR}, {\rm est.}) = N_p({\rm CR}, {\rm obs.}) \times \left [\frac{{\rm MC}_p({\rm SR}, {\rm raw})}{{\rm MC}_p({\rm CR}, {\rm raw})}\right ] = \mu_{p}\times {\rm MC}_p({\rm SR},{\rm raw}) \,,
\end{equation}
}
where $N_p({\rm SR}, {\rm est.})$ is the SR background estimate for each simulated physics processes $p$ 
considered in the analysis,
$N_p({\rm CR}, {\rm obs.})$ is the observed number of data events in the CR for the process, and ${\rm MC}_p({\rm SR}, {\rm raw})$ and ${\rm MC}_p({\rm CR}, {\rm raw})$ are raw and unnormalized estimates of the contributions from the 
process to the SR and CR respectively, as obtained from MC simulation. 
Similar equations enable the background estimates to be normalized coherently across all the CRs and the VRs.
The ratio appearing in the square brackets in Eq.~\ref{eq:tf} is defined as the transfer factor TF. 
$N_p({\rm CR}, {\rm obs})$ is often rewritten as $\mu_{p}$ multiplied by a fixed, nominal background
prediction, where $\mu_{p}$ is the actual normalization factor obtained in the fit to data.

An important feature of using TFs is that systematic uncertainties on the predicted background processes 
can be canceled in the extrapolation; a virtue of using the ratio of MC estimates.
The total uncertainty on the number of background events in the SR is then a combination of the statistical uncertainties in the CR(s) and the residual systematic uncertainties of the extrapolation. 
For this reason, CRs are often defined by somewhat looser cuts than the SR, 
in order to increase CR data event statistics without significantly increasing residual uncertainties in the TFs,
which in turn reduces the extrapolation uncertainties to the SR.
More information on the use of normalized systematic uncertainties is given in Sec.~\ref{sec:systematics}.

\section{HistFitter software framework}
\label{sec:hfframework}

HistFitter provides a programmable framework to build and test a set of data models.
To do so, HistFitter takes a user-defined configuration as input, together with raw data.
The HistFitter processing sequence then consists of three steps, illustrated by Fig.~\ref{fig:seq}.
From left to right:
\begin{enumerate}
\item Based on the user-defined configuration, HistFitter automatically prepares initial histograms, using ROOT, 
from the provided input source(s)
that model the physics processes in the data. 
(The user-defined configuration and histogram creation is discussed further below and in Sec.~\ref{sec:pdf}.)
\item According to each specified configuration, the generated histograms are combined by HistFactory to construct a corresponding PDF. 
At the end of this process, each PDF is stored in a \texttt{RooWorkspace} object together with the dataset and model configuration. 
\item The constructed PDFs are used to perform fits of the data with RooFit, perform statistical test with RooStats, 
and to produce plots and tables.
\end{enumerate}

\begin{figure}[tbh!]
\centering
\includegraphics[width=0.95\textwidth]{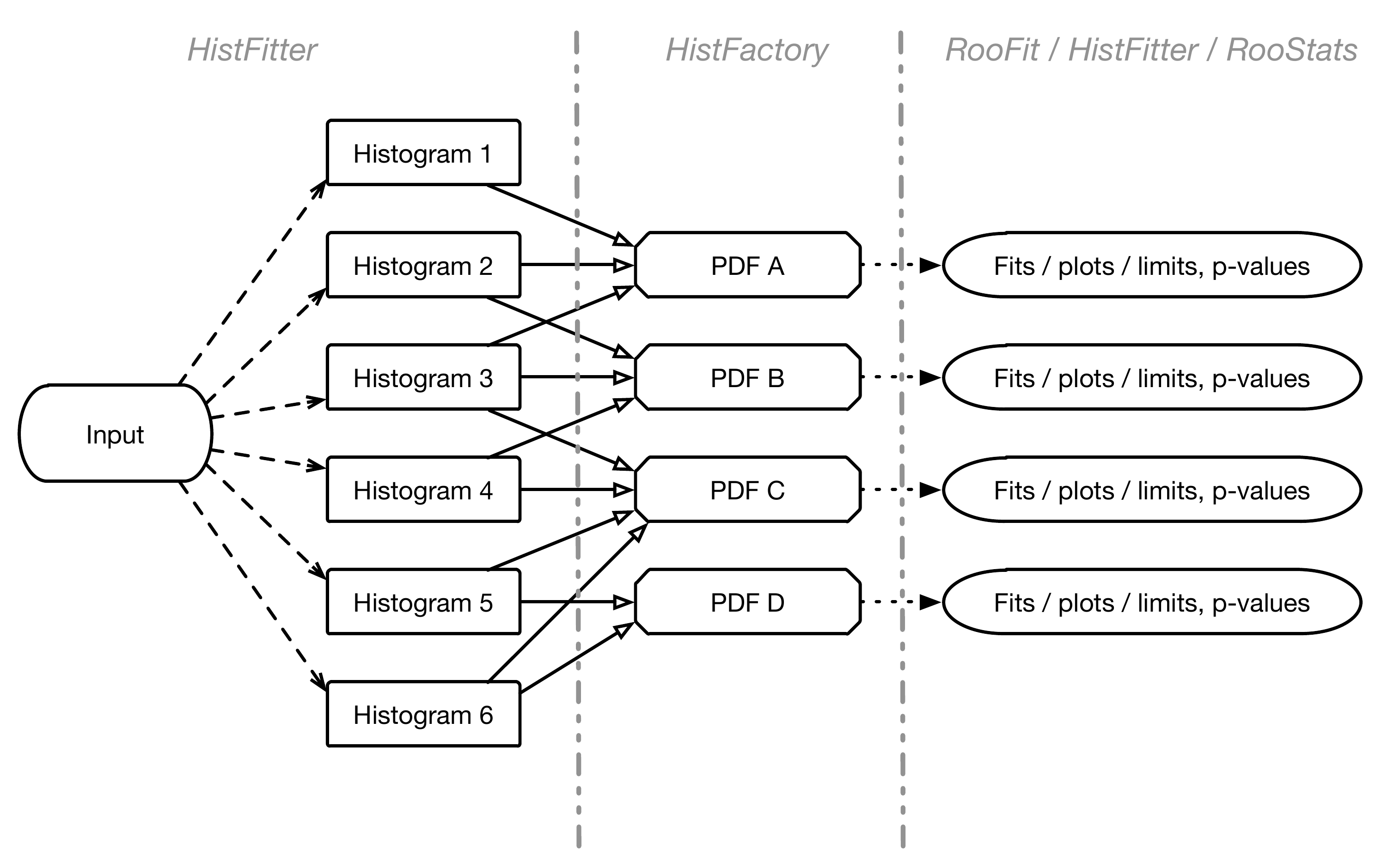}
\caption{Overview of the HistFitter processing sequence. }
\label{fig:seq}
\end{figure}

These steps all require a substantial bookkeeping and configuration machinery, which is provided by HistFitter. 
The following sub-sections summarize the central HistFitter configuration tool
and the prominent features of the HistFactory and RooStats software tools 
that HistFitter utilizes.

The various steps can can executed individually or consecutively in a single run,
and are all controlled with a single (and simple) user-defined configuration file.
For example, in early stages of an analysis cut-selections may need to be determined,
requiring frequent regeneration of just the histograms that describe the data.
Whereas, when moving to later stages in an analysis, allowing one to go straight from 
a low level description of the data 
to a high level (such as the statistical significance determination) 
can have quite a beneficial impact.

One of the key benefits of having a single configuration is to aid collaboration
between the various members of an analysis group. 
For example, the ability to rerun an analysis reasonably quickly, as long as the histograms have been created, 
helps tremendously in sharing workload between a collaboration, 
as does having the statistical tools easily accessible to the various members of an analysis group.
Likewise, the process of combining existing analyses is made more efficient than if each group has 
to independently submit histograms to some third party for a statistical combination.

\subsection{Configuration manager}
\label{sec:histfitter}
 
The central HistFitter configuration and bookkeeping machinery is built around a configuration manager,
\texttt{configManager}, implemented by two singleton objects: one in Python and one in C++. 
When executing HistFitter, users interact with the Python interface of the \texttt{configManager}
to define, for each data model, a \texttt{fitConfig} object. 
The configuration manager can hold any number of \texttt{fitConfig} objects.

A \texttt{fitConfig} object contains a PDF describing the CR, SR and VR data belonging to the model, 
together with meta-data required for the sequence of building, fitting, visualizing and interpreting 
each configuration
(\emph{i.e.} one entire row in Fig.~\ref{fig:seq}) including the generation of relevant input histograms.
The \texttt{fitConfig} class is described further in Sec.~\ref{sec:fitConfig}.

In terms of design patterns, the 
\texttt{configManager} can be seen as a ``factory of factories'',
since it generates the construction of fitConfig objects that are themselves factories of PDF objects.
By producing a \emph{list} of data models, HistFitter
introduces an additional level of abstraction which allows 
hypothesis tests to be performed 
over grids of signal models.

The construction of each data model
typically requires the preparation of tens to hundreds of histograms. 
This can lead to memory exhaustion problems for long lists of models.
However, while signal samples tend to be unique to each model, the
background samples are often identical in most of the models. 
When preparing input histograms for each sample of each model,
the configManager stores unique auto-generated names in a Python dictionary.
The dictionary is used in turn to efficiently identify and re-use the histograms
that can be shared between independent data models (see Fig.~\ref{fig:seq}),
which significantly reduces the memory usage of the software.
Additionally, the generated histograms are stored in an external file,
allowing them to be directly loaded when rerunning HistFitter in the same configuration.
This avoids the need for their, usually time-consuming, regeneration,
and helps in sharing workload between a collaboration.

\subsection{HistFactory}
\label{sec:histfactory}

HistFitter uses the HistFactory package
to construct a parametric model describing the data, based on provided input histograms.
This parametric model describes the nominal prediction and associated systematic variations  
of multiple signal and background processes in multiple regions, up to nearly arbitrary complexity.
The input histograms can be generated by HistFitter, or can be provided externally by users.

As detailed in Ref.~\cite{Cranmer:2012sba},
the PDF constructed by HistFactory describes the parameter(s) of interest, such as the rate of a signal process,
the normalization factors for background processes (as estimated from the data),
and the so-called nuisance parameters that parametrize the impact of systematic uncertainties.
Each systematic uncertainty $i$ is described with a nuisance parameter, $\theta_i$, 
that continuously interpolates between the variation and nominal templates, e.g.
$\theta_i=\pm 1$ for $\pm 1\sigma$ variations and
$\theta_i=0$ for the nominal template, where $1\sigma$ means one standard deviation.

The general likelihood $L$ of the analyses considered here is the product of 
Poisson distributions of 
event counts in the
SR(s) and/or CR(s) and of additional distributions that 
implement the constraints on systematic uncertainties.
It can be written as:
\begin{eqnarray}
\label{stat_likelihood}
  L({\boldsymbol n},\boldsymbol{\theta}^0|\mu_{\rm sig},{\boldsymbol b},{\boldsymbol\theta})&=& P_\textrm{\,SR} \;\times\; P_\textrm{\,CR} \;\times\; C_\textrm{syst}  \nonumber\\
 &=&P(n_{S}|\lambda_{S}(\mu_{\rm sig},{\boldsymbol b,\boldsymbol\theta})) 
\:\times \:
\prod_{i\in \textrm{CR}} P(n_{i}|\lambda_{i}(\mu_{\rm sig},{\boldsymbol b,\boldsymbol\theta})) 
\:\times \:
C_\textrm{syst}(\boldsymbol{\theta}^0,\boldsymbol{\theta}) \,.
\end{eqnarray}
The first two factors of Eq.~\ref{stat_likelihood} (see $P$ without subscript) reflect the Poisson measurements
of $n_{S}$ and $n_{i}$, the number of observed events in the signal
region and each control region $i$. 
The Poisson expectations $\lambda_S$ and $\lambda_i$ are functions depending on 
the predictions $\boldsymbol{b}$ for various background sources, 
the nuisance parameters that parametrize systematic uncertainties, 
the normalization factors for background processes, $\mu_{\rm p}$,
and also the signal strength parameter $\mu_{\rm sig}$.
For $\mu_{\rm sig}=0$ the signal component is turned off,
and for $\mu_{\rm sig}=1$ the signal expectation equals the nominal value of the model under consideration.

The predictions for signal and background sources are forced to be positive in HistFactory for any values of the nuisance parameters and in any histogram bin.

Systematic uncertainties are included using the probability density
function $C_\textrm{syst}(\boldsymbol{\theta}^0,\boldsymbol{\theta})$, where
$\boldsymbol{\theta}^0$ are the central values of the auxiliary measurements around which
$\boldsymbol{\theta}$ can be varied, for example when maximizing the
likelihood.  
The impact of changes in nuisance parameters on the expectation values are
described completely by the functions predicting the amount of signal and background,
$\lambda_{S}$ and $\lambda_{i}$.
For independent nuisance parameters, $C_\textrm{syst}$ is
simply a product of the probability distributions
corresponding to the auxiliary measurements describing each of the systematic uncertainties, 
typically Gaussians $G$ with unit width,
\begin{equation}
\label{syst_likelihood}
C_\textrm{syst}(\boldsymbol{\theta}^0,\boldsymbol{\theta}) = \prod_{j\in \textrm{S}} G(\theta_\textrm{j}^0-\theta_\textrm{j})\,,
\end{equation}
where S is the full set of systematic uncertainties considered.
The auxiliary measurements $\theta_\textrm{j}^0$ are typically fixed to zero, 
but can be varied when generating pseudo experiments (see below).

Several interpolation (and extrapolation) algorithms are employed in HistFactory to describe the PDF for all values
of nuisance parameters $\theta_\textrm{j}$. Some details of these algorithms are discussed in Sec.~\ref{sec:systematics},
 but for a complete overview the reader is referred to Ref.~\cite{Cranmer:2012sba}.

The execution of HistFactory results in a \texttt{RooWorkspace}, a persistent  \texttt{RooFit} object containing 
the parametrized PDF, the dataset, and a helper object summarizing the model configuration. 
As discussed in the next sub-section, these are used as input to perform statistical tests with the 
RooStats package.

\subsection{RooStats}
\label{sec:roostats}

HistFitter is capable of performing a list of pre-configured statistical tests to one or several dataset(s) 
from a single command-line call. To do so, it 
interfaces with the RooStats package. 
These tests are:
\begin{enumerate}
\item hypothesis tests of signal models;
\item the construction of expected and observed confidence intervals on model parameters. 
For example, the 95\% confidence level upper limit on the rate of a signal process;
\item the significance determination of a potentially observed event excess.
\end{enumerate}

A suite of statistical calculations can be performed, as configured by the user, 
ranging from Bayesian to Frequentist philosophies and
using various test statistic quantities as input.\footnote{
Supported test statistics are: a maximum likelihood estimate of the parameter of interest, a simple likelihood ratio
$-2\log(L(\mu,\tilde{\theta}) / L(0,\tilde{\theta}))$, as used by the LEP collaborations, a ratio of profile likelihoods $-2\log(L(\mu,\hat{\hat{\theta}}) / L(0,\hat{\theta}))$, as used by the Tevatron collaborations, or a profile likelihood ratio $-2\log(L(\mu,\hat{\hat{\theta}}) / L(\hat{\mu},\hat{\theta}))$, as used by the LHC collaborations. 
For the later case, the hypothesis tests can be evaluated as one- or two-sided. 
The sampling of the test statistics is done either with a Bayesian, Frequentist, or a hybrid calculator \cite{Moneta:2010pm}. 
}
By default, HistFitter employs a Frequentist method to perform hypothesis tests
and uses the profile likelihood ratio $q_{\mu_{\rm sig}}$ as test statistic (details below in Sec.~\ref{sec:plr}).
The CLs method~\cite{CLs} is used to test the exclusion of new physics hypothesis.
Whenever appropriate, this method is approximated by asymptotic formulae~\cite{Cowan:2010js}
to speed up the evaluation process. 
Though not strictly HistFitter specific, some details follow in the next two sub-sections.

More details about how hypothesis tests are performed with HistFitter are given in Sec.~\ref{sec:interpret}.

\subsubsection{The profile likelihood ratio}
\label{sec:plr}

As described in detail in Ref.~\cite{Cowan:2010js}, the likelihood function $L$ used in the profile likelihood ratio is built from the observed data and the parametric model that describes the data.\footnote{Note that the maximization of the likelihood function forces the need for continuous and smooth parametric models 
to describe the signal and background processes present in the data.}
The profile log likelihood ratio for one hypothesized signal rate $\mu_{\rm sig}$ is given by the test statistic:
\begin{equation}
q_{\mu_{\rm sig}} = -2 \log \bigg( \frac{L(\mu_{\rm sig},\hat{\hat{\boldsymbol\theta}})}{L(\hat{\mu}_{\rm sig},\hat{\boldsymbol\theta})} \bigg) \,,
\label{stat_lambda_eq}
\end{equation}
where $\hat{\boldsymbol \mu}_{\rm sig},\hat{\boldsymbol\theta}$ maximize the likelihood function, 
and $\hat{\hat{\boldsymbol\theta}}$ maximize the likelihood for the specific, fixed value of the signal strength $\mu_{\rm sig}$. 
Different definitions of $q_{\mu_{\rm sig}}$ apply to discovery and signal model exclusion hypothesis tests,
and also to different ranges of $\mu_{\rm sig}$, as discussed in detail in Ref.~\cite{Cowan:2010js}.

The Frequentist probability value, or $p$-value, assigned by an hypothesis test of the data, 
\emph{e.g.} a discovery or signal model exclusion test, is calculated using a distribution 
of the test statistic, $f( q_{\mu_{\rm sig}} |\, \mu_{\rm sig}, {\boldsymbol\theta})$.
This distribution can be obtained by throwing multiple pseudo experiments that randomize
the number of observed events and the central values of the auxiliary measurements.

The test statistic $q_{\mu_{\rm sig}}$ has an important property.
According to Wilks' theorem~\cite{wilks} the distribution of 
$f( q_{\mu_{\rm sig}} |\, \mu_{\rm sig}, {\boldsymbol\theta})$ 
is known in the case of a large statistics data sample. 
For a single signal parameter, $\mu_{\rm sig}$, it follows a $\chi^2$ distribution with one degree of freedom
and is independent of actual values of the auxiliary measurements, thus making it easy to
approximate. 
The case of large statistics is also called ``the asymptotic regime''.
Approximation of large statistics holds reasonably well in most cases, \emph{e.g.} from as few as $\mathcal{O}(10)$ data events.
Typically one therefore uses asymptotic formulas\footnote{Equal to taking the median and width of a collection of pseudo experiments, see discussion in Ref.\cite{Cowan:2010js}.} to evaluate the $p$-value of the hypothesis test,
avoiding the need for time-costly pseudo experiments.

\subsubsection{The profile construction}
\label{sec:profileconst}

When not working in the asymptotic regime, \emph{i.e.} in cases of low statistics, the distribution of the test statistic $f( q_{\mu_{\rm sig}} |\, \mu_{\rm sig}, {\boldsymbol\theta})$ needs to be sampled using pseudo experiments.
As the true values of the auxiliary measurement are unknown,
one ideally scans $\mu_{\rm sig}$ and ${\boldsymbol\theta}^0$ to generate
a sufficiently high number of pseudo experiments for each set. In this way one can find the values that give the 
most conservative $p$-value for the parameter of interest.
For example, one cannot exclude a signal model if there is any set of auxiliary measurement
values where the CLs value is greater than 5\%.

This is not a practical procedure when there is a large set of auxiliary measurements to consider. 
However, it turns out a good guess can be made of what values of ${\boldsymbol\theta}^0$ maximize the $p$-value.
The idea is the following. 
As a $p$-value is based on the observed data, the largest $p$-value essentially corresponds to the scenario that is most 
compatible with the data.
Therefore one first fits the nuisance parameters based on the observed data and the hypothesized value of $\mu_{\rm sig}$, 
including all fit regions. 
These are then used to set the auxiliary measurement values.
In statistics terms: the nuisance parameters have been ``profiled'' on the observed data.
Based on this, one generates the pseudo experiments that are expected to maximize the $p$-value over the auxiliary measurements,
and the observed $p$-value is evaluated as usual.
This procedure is called ``the profile construction''.

This procedure guarantees exact statistical coverage for a counting experiment
in the case where the fitted values of ${\boldsymbol\theta}^0$ correspond to their true values.
Towards the asymptotic regime, however, the distribution of 
$f( q_{\mu_{\rm sig}} |\, \mu_{\rm sig}, {\boldsymbol\theta})$ becomes independent of 
the values of the auxiliary measurements used to generate the pseudo experiments.
As a result, when using this procedure, the $p$-value obtained from the hypothesis test is robust,
and generally will not undercover.

Both the observed and expected $p$-values depend equally on the unknown true values of the auxiliary measurements.
For consistency reasons, the convention adopted at the LHC is to use the same values
to obtain the expected $p$-value as the observed $p$-value on the data. 
\emph{I.e.} the same fitted background levels are used to generate pseudo experiments for both cases,
such that the predicted expectation is the most compatible assessment for the actual observation. 
Through this choice the expected $p$-value now depends indirectly on the observed data in the SR(s).
A consequence is discussed in Sec.~\ref{sec:modindepul}.

\section{Programming of Probability Density Functions}
\label{sec:pdf}

HistFitter is designed to build and manipulate PDFs of nearly arbitrary complexity.

In the terminology of HistFactory, the likelihood function in Eq.~\ref{stat_likelihood}
has multiple \emph{channels}, 
which need inputs in the form of \emph{samples}, corresponding to the signal and background 
processes for that region. 
In turn, the various samples have systematic uncertainties, or \emph{systematics}. 
A HistFactory ``channel'' is a synonym for a ``region'', generically referring to either 
CR, SR or VR in this section.
The systematic uncertainties 
can be either statistical, theoretical or experimental in nature.
These HistFactory C++ classes are mirrored (over-loaded) by HistFitter in Python,   
and extends them by adding the flexibility to construct multiple 
PDFs from these building blocks in a programmable way, as discussed further in this section.

An example HistFitter configuration file, written in Python and demonstrating these components, 
is shown in Appendix~\ref{app:exampleconfiguration}.

\subsection{The fit configuration}
\label{sec:fitConfig}

\begin{figure}
 \begin{center}
    \includegraphics[width=\textwidth]{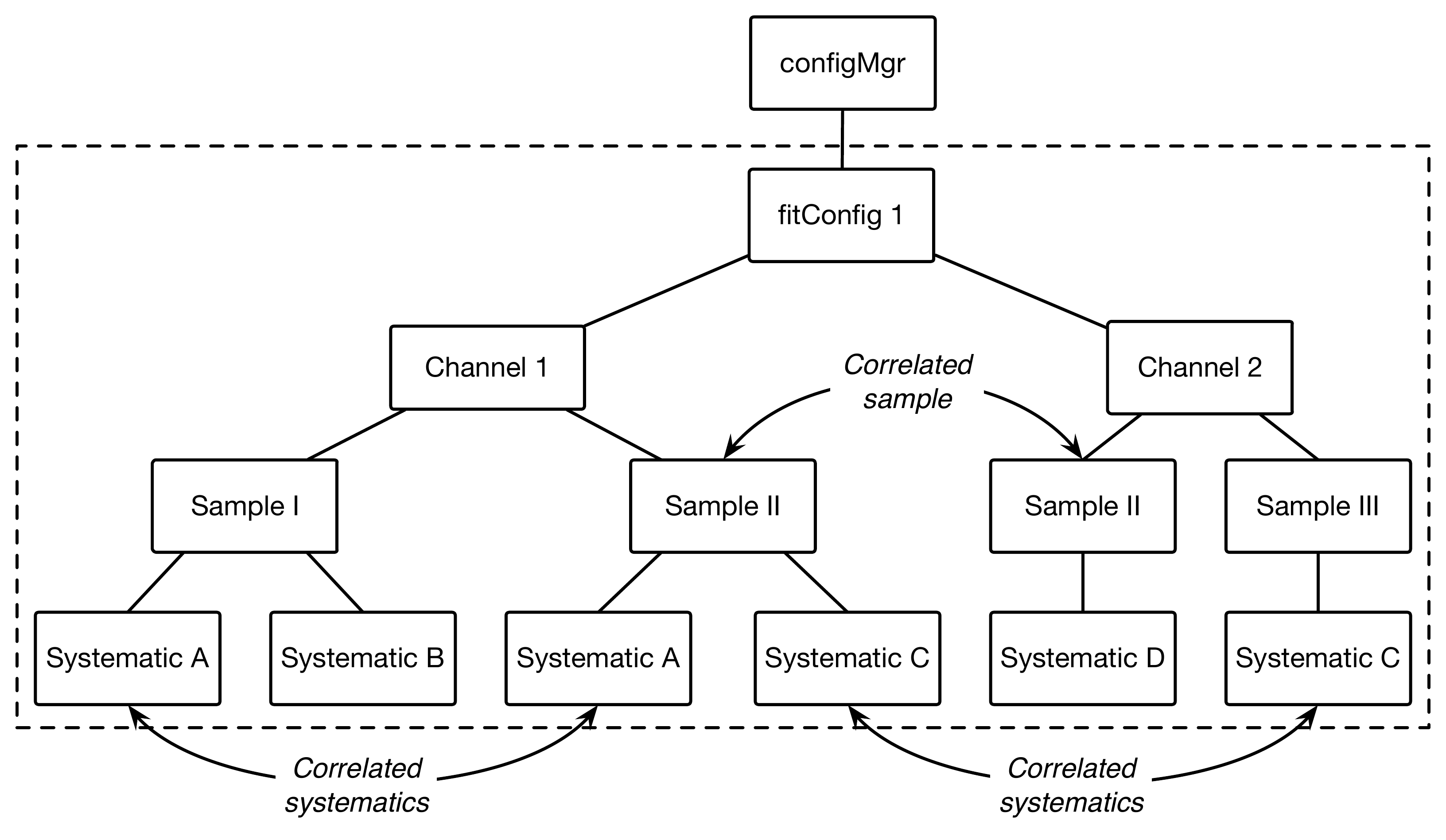}
 \end{center}
 \caption{Illustration of a fit configuration in HistFitter. Each fitConfig instance defines a PDF built from
a list of channel (\emph{i.e.} CR, SR or VR), sample and systematic objects. Each channel owns a list of
samples and each sample owns a list of systematic uncertainties. Correlated samples and systematics are declared by
being given identical names. Otherwise they are treated as un-correlated.}
 \label{fig:configMgr_singleFC}
\end{figure}

\begin{figure}
 \begin{center}
    \includegraphics[width=12cm]{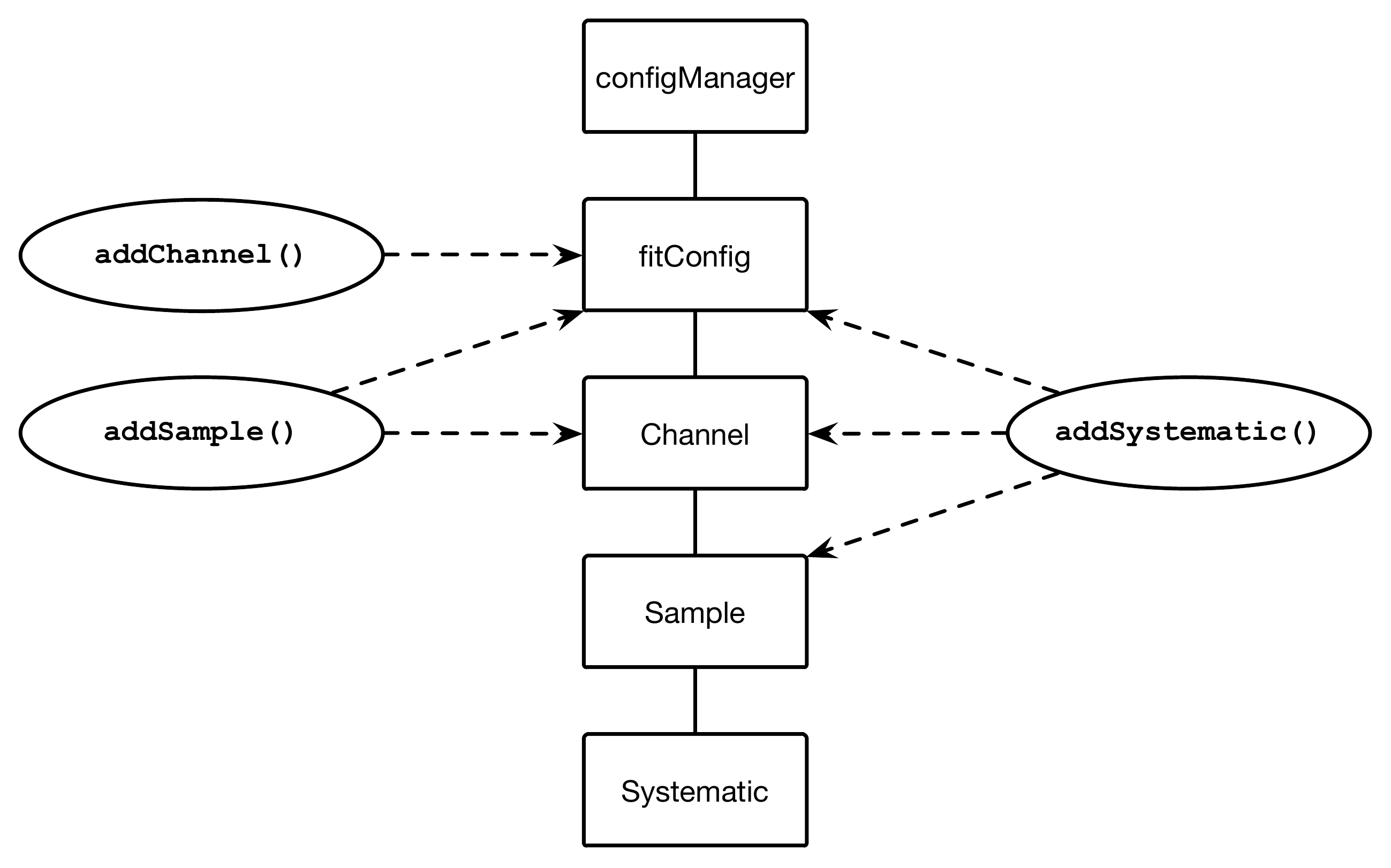}
 \end{center}
 \caption{The methods \texttt{addChannel()}, \texttt{addSample()} and \texttt{addSystematic()} are used to build 
complex PDFs in an intuitive way. The methods addSample and addSystematic implement a ``trickle down'' mechanism, discussed in the text.
}
\label{fig:configMgr_methods}
\end{figure}

HistFitter uses the \texttt{fitConfig} class to construct its PDFs. 
The design of this class
allows for the creation of highly complex PDFs, describing highly non-trivial analysis setups,
with only a few lines of intuitive code.

This is configured by users as follows:

{\scriptsize
\begin{verbatim}
from configManager import configMgr 
myFitConfig = configMgr.addFitConfig("myAnalysisName")
\end{verbatim}
}

where \texttt{myFitConfig} is a reference to a new fitConfig object owned by the configManager.
The fitConfig class logically corresponds to a PDF decorated with meta-data about
the properties of the contained channels (CR, SR, VR), including visualization, fitting and interpretation options.

During configuration, instances of channels, samples and systematics are put
together by fitConfig objects, together with links to the corresponding input histograms.
During execution, the fitConfig information is used to steer the HistFactory package's creation of 
a \texttt{RooSimultaneous} object modelling the actual PDF with RooFit.

Fig.~\ref{fig:configMgr_singleFC} illustrates the modular design of a typical HistFitter fit configuration.
The user interface provides the methods \texttt{addChannel()}, \texttt{addSample()} and \texttt{addSystematic()}
 to build up data models in an intuitive manner.
For instance, samples and systematics can be efficiently added 
to multiple channels through a ``trickle-down'' mechanism, as illustrated by Fig.~\ref{fig:configMgr_methods}.
This means that \texttt{fitConfig.addSample()} adds a sample to all the channels owned by the fitConfig, while 
\texttt{channel.addSample()} adds a sample to one specific channel. 
Similarly, \texttt{sample.addSystematic()} only adds a systematic to one specific sample while
\texttt{channel.addSystematic()} adds a systematic to all the samples owned by the channel and
\texttt{fitConfig.addSystematic()} adds a systematic to all the samples of all the channels owned by the fitConfig.

Since different channels often share the same samples (meaning: physics processes), 
and different samples often share the correlated systematic uncertainties, 
the trickle-down mechanism is in fact an extremely useful feature.
It makes it so that complex configurations of PDFs can often be described with only a few lines of code. 
As illustrated in Fig.~\ref{fig:configMgr_methods}, one simply adds all channels, samples, and 
systematic uncertainties directly to the \texttt{fitConfig} object and lets these ``trickle down'',
thereby automatically creating a highly advanced fit configuration.

A basic fit configuration can also be conveniently cloned and extended to specify new configurations,
a feature which is frequently used to build data models corresponding to multiple signal hypotheses from a common background description.

\subsection{Channels}
\label{subsec:channels}
The \texttt{Channel} objects contain data from a region of phase space defined by 
event selection criteria on the input dataset. 
Channels can represent either a simple event count (\emph{i.e.} one bin), or the multi-binned distribution of a physical observable.
New binned and un-binned channels can be added to a fitConfig by calling:

{\scriptsize
\begin{verbatim}
myChannel = myFitConfig.addChannel("myObs", ["mySelection"], nBins, varLow, varHigh) 
myUnbinnedChannel = myFitConfig.addChannel("cuts", ["mySelection"], 1, 0.5, 1.5)
\end{verbatim} 
}

where \texttt{myObs} is the name of an element of the input dataset, 
\texttt{nBins}, \texttt{varLow} and \texttt{varHigh} indicate the number of bins and the range of values
as for a one-dimensional histogram, and \texttt{mySelection} specifies the selection criteria
of the considered region.
For un-binned channels, \texttt{cuts} is a reserved keyword indicating that only the total the number 
of events passing the selection criteria needs to be considered.\footnote{This is sometimes referred to
as a ``cut-and-count'' experiment in the literature.}

As discussed in Sec. \ref{sec:fit:bkg-dis-exc}, a \texttt{Channel} object can represent a CR, SR or VR. 
This information is configured by users as follows:

{\scriptsize
\begin{verbatim}
myFitConfig.setBkgConstrainChannels(myChannel)
myFitConfig.setValidationChannels(myChannel)
myFitConfig.setSignalChannels(myChannel)
\end{verbatim}
}

It is possible to add an arbitrary number of channels to a given fitConfig by simply calling \texttt{addChannel()}
multiple times. Consequently, HistFitter automatically performs simultaneous fits constrained by the data
of all \texttt{BkgConstrainChannels} (CR) and \texttt{SignalChannels} (SR), but not by the 
\texttt{ValidationChannels} (VR). 
The data itself is described by a list of \texttt{Sample} objects owned by each channel, as discussed in the next sub-section.

\subsection{Samples}
The \texttt{Sample} class logically corresponds to a component of a RooFit PDF decorated with HistFitter meta-data. 
In a typical particle physics analysis, each sample
corresponds to a specific physics process and
several samples are needed to model a complete dataset. 

In HistFitter, samples can be defined in a specific channel or defined simultaneously in multiple channels.
The Sample class also owns a list of objects representing its systematic uncertainties.
Importantly, samples provide the link between PDF components and raw input data.
Three types of inputs are supported: 
\begin{enumerate}
\item TTree: a ROOT data structure, stored in a \texttt{TFile}, in which a list of \emph{events} 
is mapped to a list of key-value pairs characterizing the properties of each event;
\item Float: floating-point numbers provided by users through the Python interface of HistFitter;
\item Histogram: pre-made histograms using the ROOT \texttt{TH1} data structure, stored in an external \texttt{TFile}.
\end{enumerate}

The most commonly used type of input is \texttt{TTree}, which provides maximal flexibility and features but
requires the largest amount of processing power and disk I/O.
Float inputs tend to be used for quick tests and simple processes.
Histogram inputs can be used for compatibility with external frameworks, and also allow the user to
conveniently skip the \texttt{TTree}-to-histogram processing when re-building PDFs.
In all cases, the raw input is transformed into histograms as specified by Sample objects,
before being saved to a temporary file and passed to HistFactory to build the RooFit PDFs 
(see Sec. \ref{sec:fitConfig}). 

A basic sample can be created and configured by users as follows:

{\scriptsize
\begin{verbatim}
mySample = Sample("SampleName",myColor)
myChannel.addSample(mySample)
\end{verbatim}
}

which constructs a sample object owned by \texttt{myChannel}
and displayed with \texttt{myColor} color by the visualization tools. 
In this example, HistFitter takes inputs from a \texttt{TTree} object named \texttt{SampleName}
in the default ROOT file specified at the configManager level. 
To construct the sample, HistFitter uses the event selection criteria of the parent channel and
applies a default sample weight.

The default settings can be over-written by users to achieve specific goals. 
For instance, a sample can be built from Float input with:

{\scriptsize
\begin{verbatim}
mySample.buildHisto([100,34,220], "region", "observable")
\end{verbatim}
}

where the list \texttt{[100,34,220]} specifies the values of three bins in an histogram.
The default sample weight and path to the input data can also be over-written as follows:

{\scriptsize
\begin{verbatim}
mySample.setWeight(("weight1","weight2")) 
mySample.setFileList(["File1.root","File2.root"]) 
mySample.setTreeName("ArbitraryName") 
mySample.setHistoName("ArbitraryName")
\end{verbatim}
}

Weights are passed as a string to also allow the easy use of weights stored in a ROOT \texttt{TTree}. In addition, the \texttt{Sample} class has optional methods to configure its corresponding RooFit PDF, such as:

{\scriptsize
\begin{verbatim}
mySample.setStatConfig(False) 
mySample.setNormFactor("myNorm", 1.0, 0.0, 10.0)
\end{verbatim}
}

resulting in the deactivation (activated by default) of built-in Poisson statistical uncertainties,
and in the creation of a fit normalization factor \texttt{myNorm} with initial value 1.0 and 
allowed range 0.0 to 10.0, respectively.

Last but not least, HistFitter provides many features for modeling
the systematic uncertainties associated to each sample, as discussed in the next sub-section.

\subsection{Systematic uncertainties}
\label{sec:systematics}

For each model component, a nominal distribution representing the best available prediction 
is typically provided to the physics analysis as a histogram owned by a \texttt{Sample} object.
These components typically have systematic uncertainties whose
impact gets quantified in dedicated studies.
This is often modeled as variations of one standard deviation
around the nominal prediction, provided
to the physics analysis as sets of two additional histograms.
These systematic uncertainties are parametrized in the PDF with nuisance parameters, as in Eq.~\ref{stat_likelihood}.

In HistFitter, systematic uncertainties are implemented with a dedicated \texttt{Systematic}
class with several options. In a typical analysis, several \texttt{Systematic} objects are built and owned by a 
parent \texttt{Sample}.
Through the trickle down mechanism described in Sec. \ref{sec:pdf}, systematics can be defined for a specific 
sample or defined simultaneously for multiple samples and/or multiple channels. 

A \texttt{Systematic} object can be conceived as a doublet of samples
specifying up and down variations around the parent \texttt{Sample}. 
Hence \texttt{Systematic} objects can be constructed from the same types of inputs as \texttt{Samples}, 
namely: \texttt{TTree}, Float and histogram. 

When using \texttt{TTree} inputs, two methods can be used to compute
the up/down variations of a systematic: weight-based or tree-based. In the weight-based method,
histograms are always built from the same \texttt{TTree}, using three different sets of weights: 
up, nominal and down. In the tree-based method, 
histograms are built from three different \texttt{TTree}s using the same set of weights.
If only one variation is available, users can either build a one-sided uncertainty or 
symmetrize the variation as $\mathrm{\frac{nominal\pm(up-nominal)}{nominal}}$.

\texttt{Systematic} objects can be created by users as follows:

{\scriptsize
\begin{verbatim}
mySys = Systematic("myTreeSys", "ASample", "ASample\_UP", "ASample\_DOWN", "tree", "myMethods") 
mySys = Systematic("myWeightSys", ["nominalWeights"], ["upWeights"], ["downWeights"], "weight", "myMethods") 
mySys = Systematic("myUserSys", ["nominalWeights"], 1.1, 0.8, "user", "myMethods")
\end{verbatim}
}

where \texttt{myTreeSys} and \texttt{myWeightSys} rely on the tree-based and weight-based methods.
\texttt{myUserSys} relies on the Float input discussed above, and, in this example,
has asymmetric up and down input uncertainty values of 10\% and 20\%.
The last argument \texttt{myMethods} is discussed below.
Systematic objects are then associated to \texttt{Sample} or \texttt{Channel} objects with: 

{\scriptsize
\begin{verbatim}
mySample.addSystematic(mySys)
myChannel.addSystematic(mySys)
\end{verbatim}
}

As illustrated in Fig.~\ref{fig:configMgr_singleFC}, 
correlated systematic uncertainties are declared simply by giving them identical names
in the corresponding \texttt{Sample}s. Otherwise they are treated as uncorrelated.

\begin{table}[t!]
\begin{center}
\begin{tabular}{|l|p{0.7\textwidth}|}
\hline
\multicolumn{2}{|l|}{Basic systematic methods in HistFactory} \\
\hline
\texttt{overallSys} & uncertainty of the global normalization, not affecting the shape \\
\texttt{histoSys} & correlated uncertainty of shape and normalization \\
\texttt{shapeSys} & uncertainty of statistical nature applied to a sum of samples, bin by bin \\
\hline
\multicolumn{2}{|l|}{Additional systematic methods in HistFitter} \\
\hline
\texttt{overallNormSys} & overallSys constrained to conserve total event count in a list of region(s)\\ 
\texttt{normHistoSys} & histoSys constrained to conserve total event count in a list of region(s) \\
\texttt{normHistoSysOneSide} & one-sided normHistoSys uncertainty built from tree-based or weight-based inputs\\
\texttt{normHistoSysOneSideSym} & symmetrized normHistoSysOneSide \\
\texttt{overallHistoSys} & factorized normalization shape and uncertainty, described with \texttt{overallSys} and \texttt{histoSys} respectively \\
\texttt{overallNormHistoSys} & overallHistoSys in which the shape uncertainty is modeled with a normHistoSys
and the global normalization uncertainty is modeled with an overallSys\\
\texttt{shapeStat} & shapeSys applied to an individual sample\\
\hline
\end{tabular}
\caption{Sub-set of the systematic methods available in HistFitter.
The methods are specified by a string argument containing a combination of basic HistFactory methods 
and optional HistFitter keywords: \texttt{norm}, \texttt{OneSide} and/or \texttt{Sym}.
Systematic objects can be built with Tree-based, weight-based, Float or histogram input methods
in all cases.}
\label{tab:systematics}
\end{center}
\end{table}

When turning the above into nuisance parameters,
additional input is required to specify the interpolation (extrapolation) algorithm and constraint parametrization 
for each systematic uncertainty.
This is done with the argument \texttt{myMethods} above.
Several possible analysis strategies can be envisaged, requiring a detailed discussion, case by case. 
To address this, HistFitter does not enforce a specific strategy but provides users with as many methods
as possible to cover all reasonable possibilities. 

The basic methods for systematic uncertainties defined in HistFactory are called: \texttt{overallSys}, \texttt{histoSys} 
and \texttt{shapeSys}, and are listed in the top half of Tab.~\ref{tab:systematics}.

An \texttt{overallSys} describes an uncertainty of the global normalization of the sample. 
This method does not affect the shape of the distributions of the sample. 
A \texttt{histoSys} describes a correlated uncertainty of the shape and normalization. 
Both methods use a Gaussian constraint to model an uncertainty
and allow for asymmetric uncertainties, simply by providing asymmetric input values or histograms, respectively. 
By default they are configured to use a 6th-order polynomial interpolation technique between the $\pm 1 \sigma$ and nominal
histograms and a linear extrapolation beyond $|1\sigma|$~\cite{Cranmer:2012sba},
though they can be configured differently.

A \texttt{shapeSys} describes an uncertainty of statistical nature, typically arising from limited MC statistics. 
In HistFactory, \texttt{shapeSys} is modeled with an independent parameter for each bin of each channel that is, however,
shared between all samples with \texttt{StatConfig==True} (see Sec.~\ref{subsec:channels}).
For simplicity, users can also set a threshold below which samples are neglected when building a \texttt{shapeSys}.
The interpolation and extrapolation technique used for \texttt{shapeSys} are as for \texttt{histoSys}, 
and parametrized as a Poissonian constraint.

To respond to various use cases encountered during real-life analysis of ATLAS Run-1 data,
HistFitter provides additional systematic methods derived from the basic HistFactory methods.
A sub-set of the systematic methods available in HistFitter is listed in the bottom half of Tab.~\ref{tab:systematics}.

These methods can be specified with combinations of the \texttt{norm}, \texttt{OneSide} and \texttt{Sym}
keywords. 
The \texttt{norm} keyword indicates that the total event count is required to remain invariant in 
a user-specified list of normalization region(s) when constructing up/down variations.
This describes uncertainties of the shape only.
Such a systematic uncertainty
is transformed from an uncertainty on event counts in each region into a systematic uncertainty on the
transfer factors, as discussed in Sec.~\ref{sec:analysisStrategy} (Eq.~\ref{eq:tf}).  
The \texttt{OneSide} and \texttt{Sym} keywords indicate that a one-sided or a symmetrized uncertainty
should be constructed when using tree-based or weight-based inputs.

\section{Performing fits}
\label{sec:fit}

Different fit strategies are commonly used in physics analyses, 
differing by the usage of particular combinations of CRs, SRs, and VRs,
and by the consideration of a signal model or not. 
Fit strategies aim to either derive background estimates in VRs and SRs, 
or to make quantitative statements on the compatibility of the background estimate(s) 
with the observed data in the SR(s).
HistFitter is tailored specifically to the design and implementation of such fit strategies.

Also discussed in this section are the technical and statistical details 
of the extrapolation of background processes across CRs, 
SRs and VRs.

\subsection{Common fit strategies} 
\label{sec:fit:bkg-dis-exc}

The three most commonly used fit strategies in HistFitter are defined as: 
the ``background-only fit''; the ``model-dependent signal fit''; and the ``model-\emph{in}dependent signal fit''\footnote{Other 
nomenclature (but deemed confusing) for the model-dependent and model-independent signal fit are ``exclusion fit'' and ``discovery fit'' respectively.}.
This section describes the details of each fit strategy, as also
summarized in Tab.~\ref{table:fitSetups} at the end of the section.
The application of these to validation and hypothesis-testing purposes is described in detail 
in Secs.~\ref{sec:pres} and~\ref{sec:interpret}, respectively.

\subsubsection*{Background-only fit}

The purpose of this fit strategy is to estimate the total background in SRs 
and VRs, without making assumptions on any signal model. 
As the name suggests, only background samples are used in the model.
The CRs are assumed to be free of signal contamination.
The fit is only performed in the CR(s), and the dominant background processes 
are normalized to the observed event counts in these regions.
As the background parameters of the PDF are shared in all different regions,
the result of this fit is used to predict the background event levels in the SRs and VRs.

The background predictions from the background-only fit are independent of the observed 
number of events in each SR and VR, as only the CR(s) are used in the fit.
This allows for an unbiased comparison between the predicted and observed number of events in each region.
In Sec.~\ref{sec:pres} the background-only fit predictions are used to present 
the validation of the transfer factor-based background level predictions.

Another important use case for background-only fit results in the SR(s) is for external groups
to perform an hypothesis test on an untested signal model, which has not been studied by the experiment. 
With the complex fits currently performed at the LHC, it may be difficult (if not impossible) for
outsiders to reconstruct these. An independent background estimate in the SR,
as provided by the background-only fit, is then the correct estimate
to use as input to any hypothesis test (see Sec.~\ref{sec:interpret}).

Technical details of the extrapolation approach are discussed in Sec.~\ref{sec:errorpropagation},
and validation examples are given in Sec.~\ref{sec:pres}.

\subsubsection*{Model-dependent signal fit}

This fit strategy is used with the objective of studying a specific signal model. 
In the absence of a significant event excess in the SR(s), 
as concluded with the background-only fit configuration, exclusion limits can be set on the signal models under study. 
In case of excess, the model-dependent signal fit can be used to measure properties such as the signal strength.
The fit is performed in the CRs and SRs simultaneously. 
Along with the background samples, a signal sample is included in all regions, not just the SR(s),
to correctly account for possible signal contamination in the CRs.
A normalization factor, the signal strength parameter $\mu_{\rm sig}$, is assigned to the signal sample.

Note that this fit strategy can be run with multiple SRs (and CRs) 
simultaneously,
as long as these are statistically independent, non-overlapping regions. 
If multiple SRs are sensitive to the same signal model, 
performing the model-dependent signal fit on the statistical combination of these regions
shall, in general, give better (or equal) exclusion sensitivity than 
obtained in the individual analyses. 
An example of this is given in Fig.~\ref{fig:interpret:contour} (right) of Sec.~\ref{signalhypotest}.
As shown in Sec.~\ref{sec:pdf}, combining the channels of multiple analyses into 
a single fit configuration is a straightforward exercise in HistFitter.

In a similar fashion, using multiple bins of a signal-sensitive observable in the 
definition of the SR(s) will generally give a better sensitivity to any signal model 
studied\footnote{Both the addition of simultaneous SRs and of shape information in these SRs will 
make an analysis more versatile.
Since the shape of signals models over multiple bins or multiple SRs will in general be different
from the background prediction, in doing so the fit has gained separation power to distinguish the two.
In particular, the sensitivity to signal models not considered in the optimization of the SRs is often retained. 
}.
An example of such a ``shape-fit'' signal region is shown in Fig.~\ref{fig:SRs_2LSS} of Sec.~\ref{sec:visualization}.

Typically, a grid of signal samples for a particular signal model is produced
by varying some of its model parameters, such as the masses of supersymmetric particles.
The model-dependent signal fit is repeated for each of these grid points, 
thereby probing the phase space of the model.
Examples of this are provided in Secs.~\ref{signalhypotest} and~\ref{signalupperlimit}.

\subsubsection*{Model-independent signal fit}

An analysis searching for new physics phenomena 
typically sets model-independent upper limits on 
the number of events beyond the expected number of events in each SR. 
In this way, for any signal model of interest, anyone can estimate the number 
of signal events predicted in a particular signal region
and check if the model has been excluded by current measurements or not.

Setting the upper limit is accomplished by performing a model-independent signal fit. 
For this fit strategy, both the CRs and SRs are used, in the same manner as for the model-dependent signal fit.
Signal contamination is not allowed in the CRs, but
no other assumptions are made for the signal model, also called a ``dummy signal'' prediction.
The SR in this fit configuration is constructed as a single-bin region,
since having more bins requires assumptions on the signal spread over these bins. 
The number of signal events in the signal region is added as a parameter to the fit. 
Otherwise, the fit proceeds in the same way as the model-dependent signal fit.

Examples of upper limits on numbers of beyond the SM events, obtained using this setup, 
are provided in Sec.~\ref{sec:modindepul}.

The model-independent signal fit strategy, fitting both the CRs and each SR, is also used 
to perform the background-only hypothesis test, which quantifies 
the significance of any observed excess of events in a SR, again in a 
manner that is independent of any particular signal model.
More details on the background-only hypothesis test are discussed in Sec.~\ref{sec:discpvalue}.
One main but subtle difference between 
the model-independent signal hypothesis test (Sec.~\ref{sec:modindepul}) 
and 
the background-only hypothesis test (Sec.~\ref{sec:discpvalue})
is that the signal strength parameter is set to
one or zero 
in the profile likelihood numerator respectively.

\begin{table}[t!]
\begin{center}
{\small
\begin{tabular}{|l|c|c|c|}
\hline
{\bf Fit setup}           & \emph{Background-only fit}    &   \emph{Model-dependent}        &\emph{Model-independent} \\
                             &                           & \emph{signal fit}             & \emph{signal fit}    \\
\hline
{\bf Samples used}           & backgrounds &   backgrounds + signal       &  backgrounds +    \\
                             &                 &                          & dummy signal      \\
\hline
{\bf Fit regions}           & CR(s) &   CR(s) + SR(s)        & CR(s) + SR   \\
\hline
\end{tabular}
}
\end{center}
\caption{Summary of the fit strategies supported in HistFitter, as described in the text.}
\label{table:fitSetups}
\end{table}

\subsection{Extrapolation and error propagation}
\label{sec:errorpropagation}

This section discusses the extrapolation of coherently normalized background estimates from the CR(s) to each SR and VR,
as obtained from the background-only fit\footnote{The background-only fit estimates are sometimes called ``blinded'' background 
estimates, as the SR(s) and VR(s) are not included in the fit.}, discussed in Sec.~\ref{sec:fit:bkg-dis-exc}.
The basic strategy behind the background extrapolation approach is to share the background parameters 
of the PDF in all the different regions: CRs, SRs and VRs.

As discussed in Sec.~\ref{sec:histfactory}, a likelihood function 
is built from both the parametric model and the observed data.
In other words, a background-only fit to the CRs technically requires a PDF modeling only the CRs.
On the other hand, the extrapolation of the normalized background processes from the CRs to the SRs and VRs,
which uses the background-only fit result, requires a \emph{different} PDF containing all these regions.

In HistFitter, the technical construction of these various PDFs proceeds as follows.
First a total PDF describing all CRs, SRs and VRs is constructed using HistFactory.
This PDF is not used to fit the data, as the likelihood is unaware of the concept of different region types. 
HistFitter has dedicated functions to deconstruct and reconstruct PDFs, based on the channel types discussed in Sec.~\ref{subsec:channels}.

To perform the background-only fit, the total PDF is deconstructed and then reconstructed describing only the CRs. 
The result of the background-only fit is stored, containing the values, the errors and the covariance matrix corresponding to all fit parameters.
After this fit, the normalized backgrounds are extrapolated to the SRs (or VRs).
For this HistFitter deconstructs and reconstructs the total PDF, now describing the CRs and SRs (or VRs).
The background-only fit result is then incorporated into this PDF to obtain the extrapolated background prediction $b$
in any SR (or VR).

Once the background-only fit to data has been performed and the total PDF been reconstructed, 
an estimate of the uncertainty on an extrapolated background prediction $\sigma_{b,\,tot}$ can be calculated.
The determination of this error requires the uncertainties and correlations from the stored fit result.
The total error on $b$ is calculated using the typical error propagation formula
\begin{equation}
\sigma_{b,\,tot}^2 = \sum\limits_{i}^n \bigg(\frac{\partial b}{\partial \eta_i}\bigg)^2 \sigma^2_{\eta_i} + \sum\limits_{i}^n \sum\limits_{j\neq i}^n \rho_{ij} \bigg(\frac{\partial b}{\partial \eta_i}\bigg) \bigg(\frac{\partial b}{\partial \eta_j}\bigg) \sigma_{ \eta_i}\sigma_{ \eta_j}\,,
\label{eq:errorProp}
\end{equation}
where $\eta_{i}$ are the floating fit parameters, consisting of normalization factors $\mu_k$ and nuisance parameters $\theta_l$, $\rho_{ij}$ is the correlation coefficient, between $\eta_{i}$ and $\eta_{j}$, and $\sigma_{\eta_i}$ is the standard deviation of $\eta_i$. 
Any partial derivatives to $b$ are evaluated on the fly.

The after-fit parameter values, errors and correlations are saved in the RooFit class \texttt{RooFitResult}. 
Let us take an example of a background-only fit result (from CRs only) that needs to be extrapolated to a SR.
The total PDF (consisting of CRs and SRs) contains a set of parameters that can be subdivided as follows. 
\begin{enumerate}
\item There is a large set of parameters shared between CRs and SR, ${\boldsymbol \eta}_{\mathrm{shared}}$,
for example the background normalization factors and most systematic uncertainties.  
\item There is a subset of parameters connected only to the CRs, ${\boldsymbol \eta}_{\mathrm{CR}}$,
for example the uncertainties due to limited Monte Carlo statistics in the CRs.
\item And finally, there is another subset of parameters connected only to the SR, ${\boldsymbol \eta}_{\mathrm{SR}}$.
\end{enumerate}
When the fit is performed with the (deconstructed) CRs-only PDF, only the parameters ${\boldsymbol \eta}_{\mathrm{shared}}$ and 
${\boldsymbol \eta}_{\mathrm{CR}}$  are evaluated and saved in the fit result.

Hence when this fit result is propagated to the SR, the estimated error only contains 
the parameters that are shared between the CRs and the SR, and thus is incomplete. 
The uncertainties corresponding to ${\boldsymbol \eta}_{\mathrm{SR}}$ are 
not picked up in Eq.~\ref{eq:errorProp}, as these are not contained in the fit result.

HistFitter uses an expanded version of the \texttt{RooFitResult} class, called \texttt{RooExpandedFitResult}, 
that contains all of the nuisance parameters of all regions in the extrapolation PDF, 
even if these are not used in the background-only fit configuration.
This expansion makes it possible to extrapolate all of the shared parameters, while keeping the unshared parameters, 
such that a complete error can be calculated in any region. 
In the expanded fit result, the correlations between the shared and unshared parameters are set to zero.

Using the \texttt{RooExpandedFitResult} 
class the VRs can now provide a rigorous statistical cross check. 
If the background-only fit to the CRs finds that changing the background normalization and/or 
shape parameters of a kinematic distribution
gives a better description of the data, this will be reflected automatically in the VRs. 
Likewise, if the uncertainty on these parameters has a strong impact, and is reduced by the fit, 
the effect will be readily propagated.

In HistFitter, the before-fit parameter values, errors and correlations are stored in an expanded fit result object
as well.
The before-fit and after-fit background value and uncertainty predictions can thus be easily compared.
A few assumptions are made to construct this before-fit object. 
First, all correlations are set to zero prior to the fit, effectively taking out the second term of the Eq.~\ref{eq:errorProp}. 
Second, the errors on the normalization factors of the background processes are unknown prior to the fit, 
and hence set to zero. 

\begin{figure}[tp]
\begin{center}
    \includegraphics[width=0.6\textwidth]{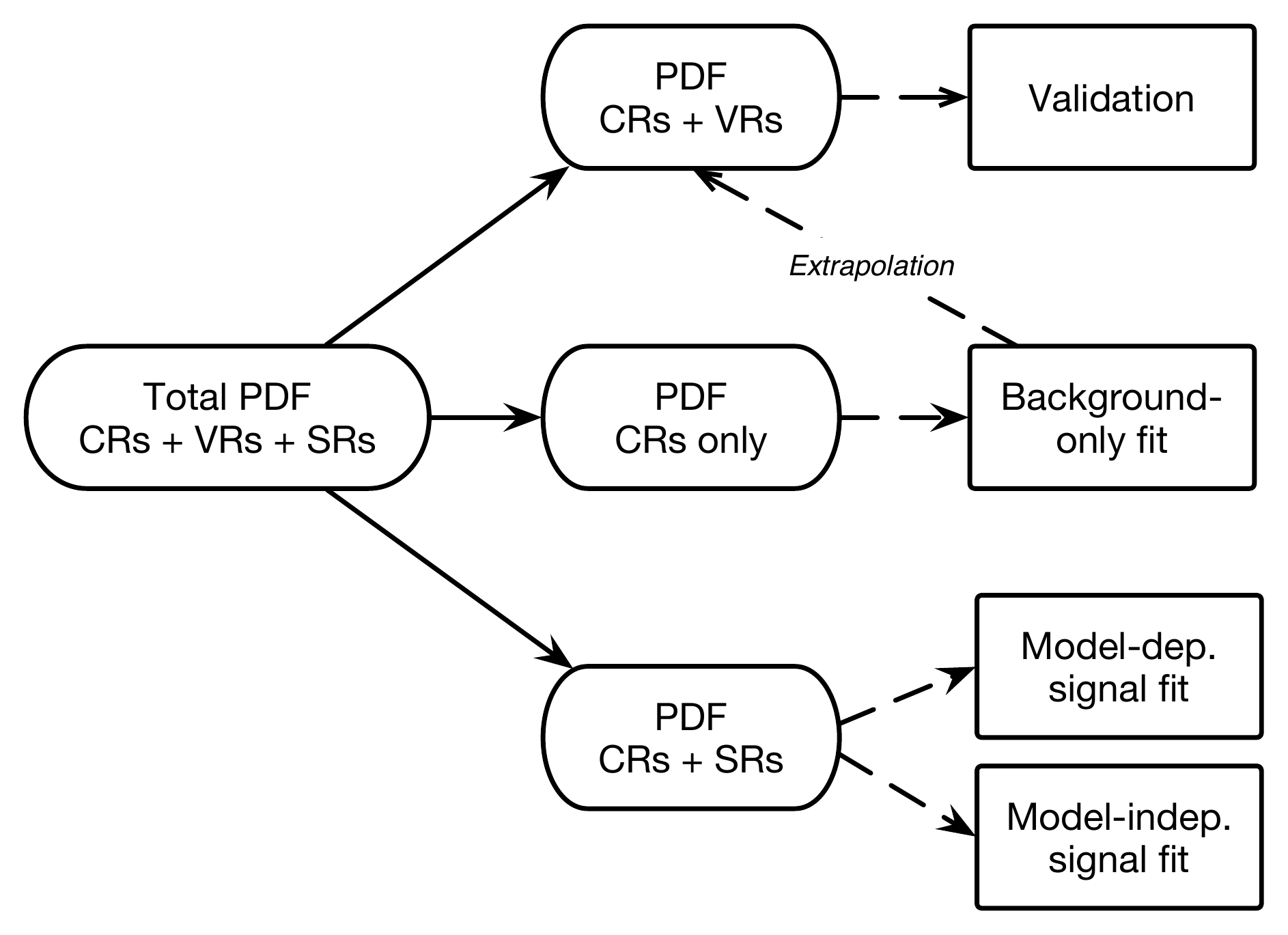}
    \caption{An overview of the various PDFs HistFitter uses internally, together with their typical use. 
The large PDF for all regions is automatically deconstructed into separate, smaller ones defined on those subsets of regions depending on the fit and/or statistical test performed. 
The PDFs are indicated as rounded squares, and the fit configurations as squares. 
}
    \label{fig:pdf_structure}
\end{center}
\end{figure}

The fit strategies of Sec.~\ref{sec:fit:bkg-dis-exc} are illustrated in Fig.~\ref{fig:pdf_structure},
together with the PDF restructuring detailed in this section.
In Fig.~\ref{fig:pdf_structure}, the various constructed PDFs are indicated as rounded squares and 
the fit configurations, on the right-hand side, as squares.

\section{Presentation of results}
\label{sec:pres}

HistFitter also contains an extensive array 
of user-friendly functions and scripts, which help with the understanding and detailing of the results obtained from the fits. 
These scripts and plotting functions are generalized, such that for every model built with HistFitter all of 
these features come without any need for further coding. 
All scripts and plotting functions can be called by single-line commands.

Two main presentation components are the visualization of fit results and scripts for producing event yield and uncertainty tables.
Both rely critically on the fits to data and uncertainty extrapolation features
discussed in Sec.~\ref{sec:fit}. 
All tables and plots, discussed in the next two sections, can be produced for any fit configuration 
of a defined model, as well as before and after the fit to the data. 
Multiple details, such as the legends on plots or the set of regions to be processed for tables, 
can easily be set in the configuration file or from the command line.

All tables and figures shown in this section come directly from publications 
by the ATLAS collaboration, 
and serve only as illustrations of the HistFitter tools that are discussed.

\subsection{Visualization of fit results}
\label{sec:visualization}

HistFitter can produce several classes of figures to visualize fit results, as detailed below.

Fig.~\ref{fig:channel_1L} shows an example plot of a multi-bin (control) region 
before (left) and after (right) the fit to the data, taken from Ref.~\cite{PhysRevD.86.092002}. 
Similar plots can be produced for any region defined in HistFitter, either single- or multi-binned.
Each sample in the example region (channel) is portrayed by a different color.
The samples can also be plotted separately (not shown), 
for the purpose of understanding the distribution of the uncertainties over the samples. 
The impact of the fit to data can be studied by comparing the before-fit to after-fit distributions. 
In the after-fit plot, as a result of the fit, the normalization, shapes and corresponding uncertainties of 
the background samples have been adjusted to best describe the observed data over all bins\footnote{Realize that one single-bin control region only has a handle on the normalization of one background sample, and not on the background shape.},
illustrating the analysis strategy outlined in Sec.~\ref{sec:analysisStrategy}.

An example of two multi-binned SRs, used in Ref.~\cite{Aad:2014pda}, is 
shown in Fig.~\ref{fig:SRs_2LSS}. 
Two different supersymmetry models, for which each SR is sensitive to strong variations over the bins, are superimposed on the before-fit background predictions.
%

\begin{figure}[h!]
\centering
\includegraphics[width=0.485\textwidth]{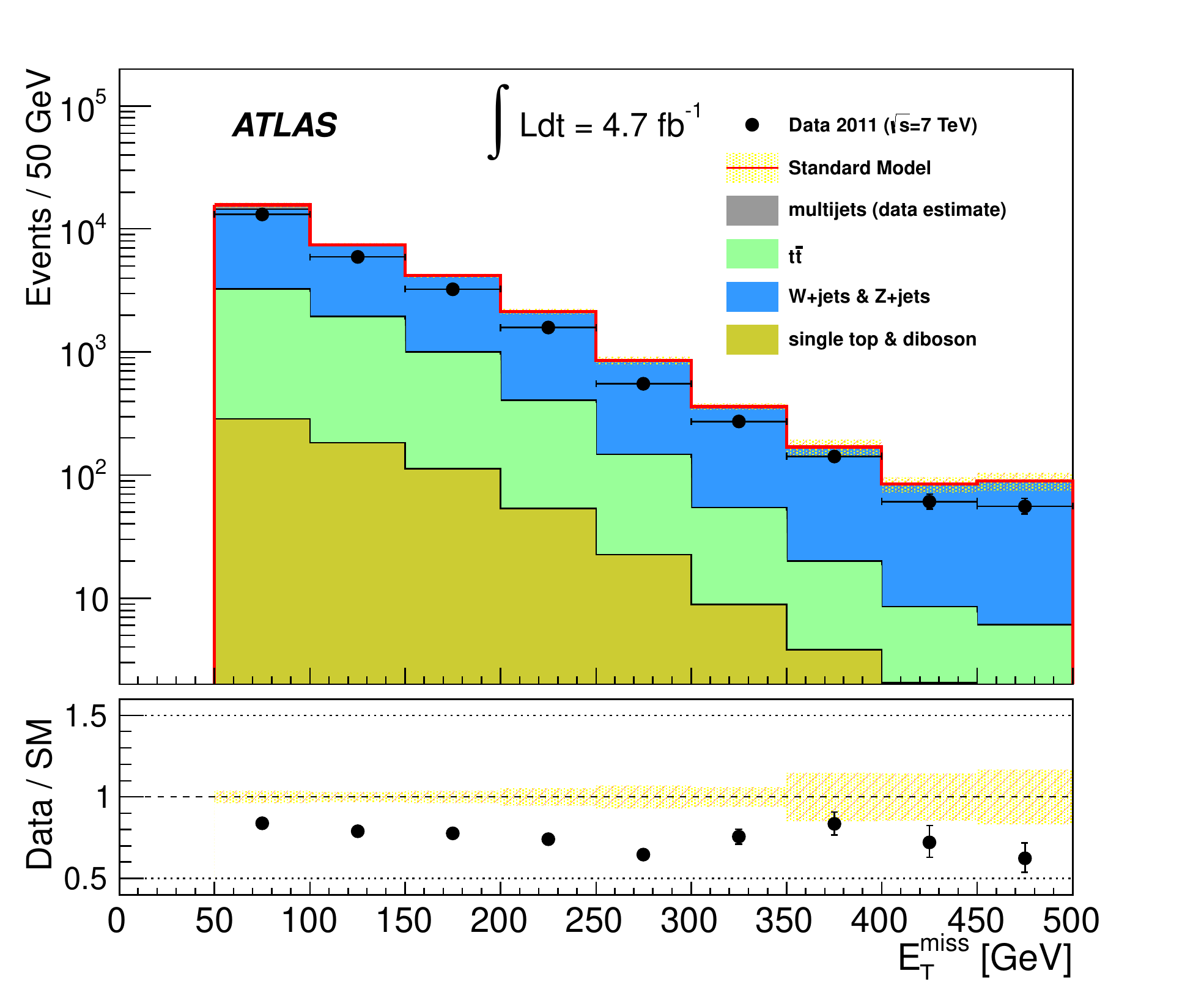}
\includegraphics[width=0.485\textwidth]{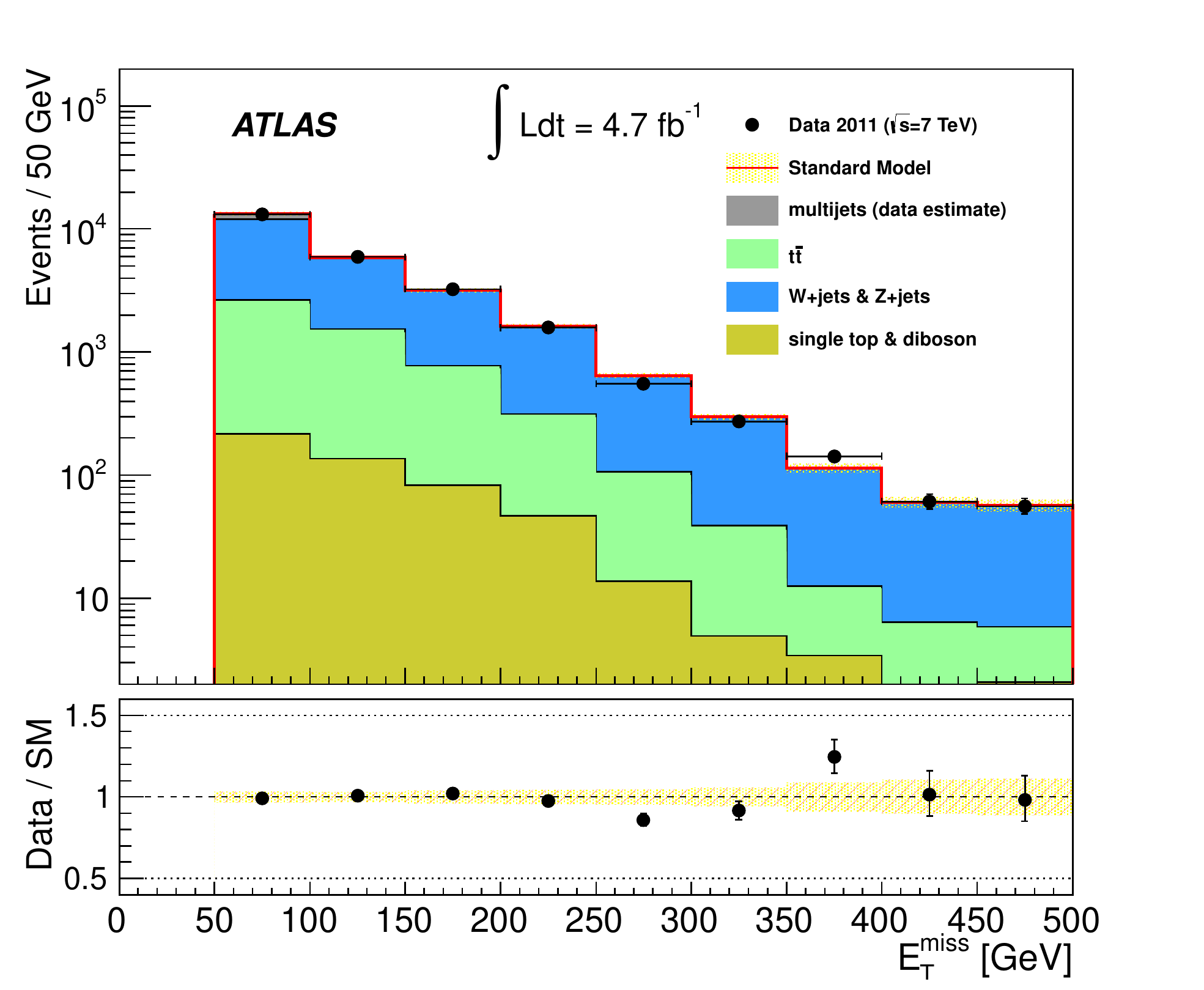}
\caption{Example produced by the ATLAS collaboration and taken from Ref.~\cite{PhysRevD.86.092002}. Distribution of missing transverse momentum in the single lepton $W$+jets control region before (left) and after (right) the final fit to all background control regions. }
\label{fig:channel_1L}
\end{figure}

\begin{figure}[h!]
\centering
{\hspace{-0.15cm}}
\includegraphics[width=0.455\textwidth]{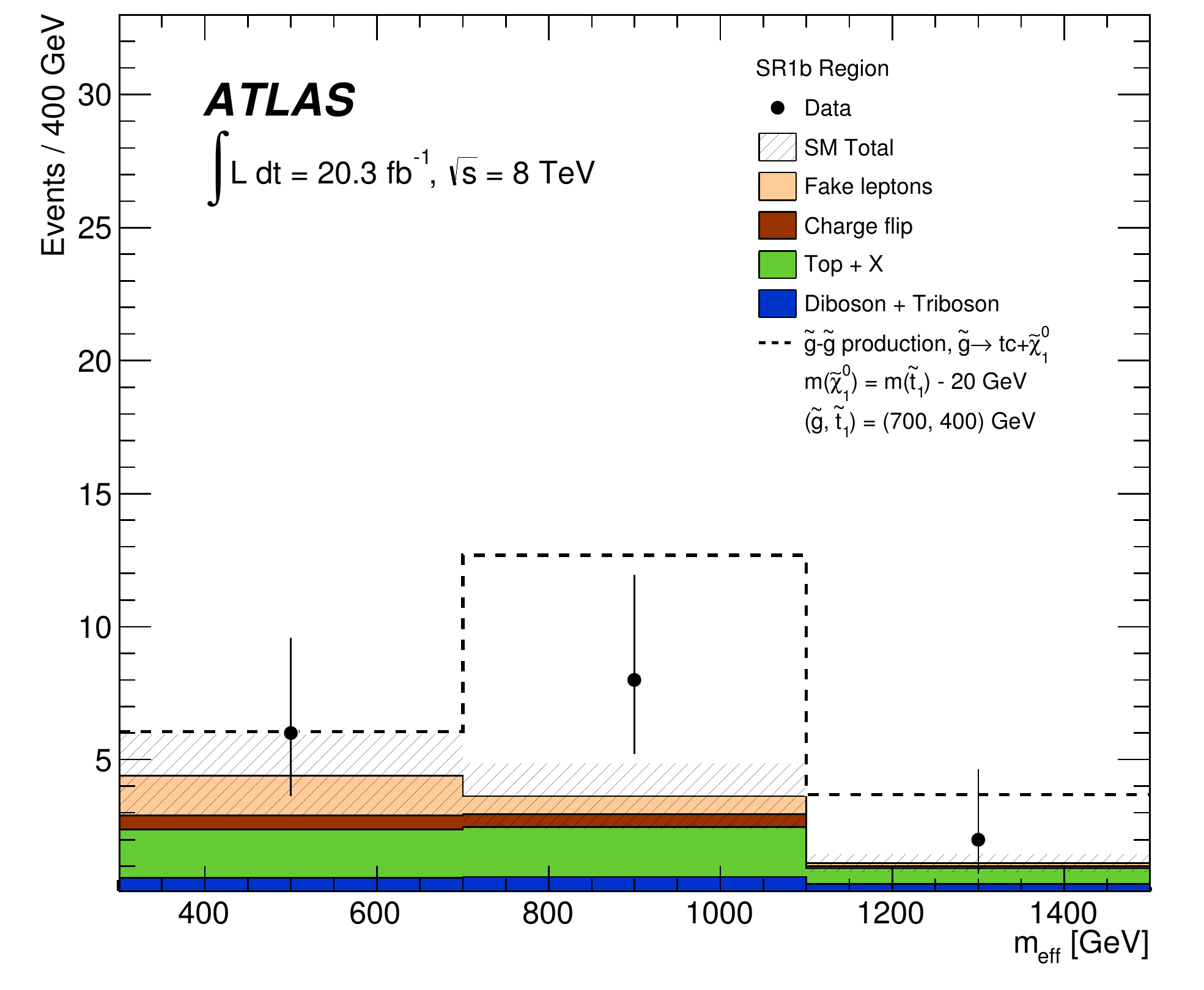} 
{\hspace{0.25cm}}
\includegraphics[width=0.455\textwidth]{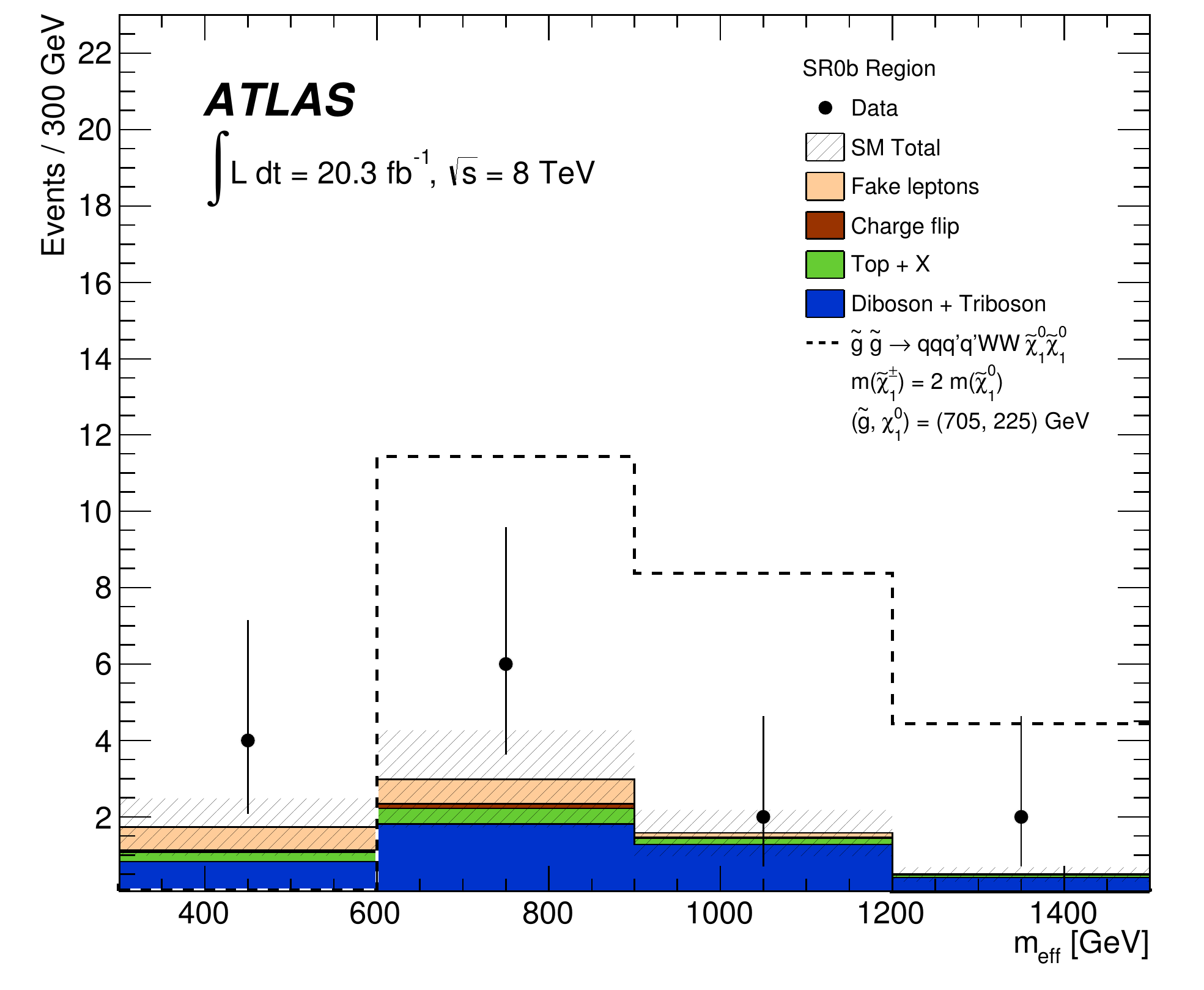}
{\hspace{0.15cm}}
\caption{Example produced by the ATLAS collaboration and taken from Ref.~\cite{Aad:2014pda}. Effective mass distributions in the signal regions SR0b and SR1b, used as input for model-dependent signal fits. For illustration, predictions are shown from two SUSY signal models with particular sensitivity in each signal region.}
\label{fig:SRs_2LSS}
\end{figure}

Fig.~\ref{fig:pull_1L} shows an example of a \emph{pull} distribution for a set of 
non-overlapping VRs, as produced with HistFitter and taken from Ref.~\cite{PhysRevD.86.092002}.
The example relies on the background prediction in each VR, as obtained from a fit to the CRs,
and tests the validity of the transfer-factor based extrapolation.
The pull $\chi$ is calculated as the difference between the observed $n_{\rm obs}$ and predicted event numbers $n_{\rm pred}$, 
divided by the total systematic uncertainty on the background prediction, $\sigma_{\rm pred}$, 
added in quadrature to the Poissonian variation on the expected number of background events, $\sigma_{\rm stat,\,exp}$.
\begin{eqnarray}
\chi &=& \frac{n_{\rm obs} - n_{\rm pred}}{\sigma_{\rm tot}} \\
\sigma_{\rm tot} &=& \sqrt{\sigma_{\rm pred}^2 + \sigma_{\rm stat,\,exp}^2}
\end{eqnarray}
If, on average, the pulls for all the validation regions would be negative (positive), the data is 
overestimated (underestimated) and the model needs to be corrected.
If the background model is properly tuned, 
on average good agreement is found between the data and the estimated background model.

\begin{figure}[h!]
\centering
\includegraphics[width=0.35\textwidth]{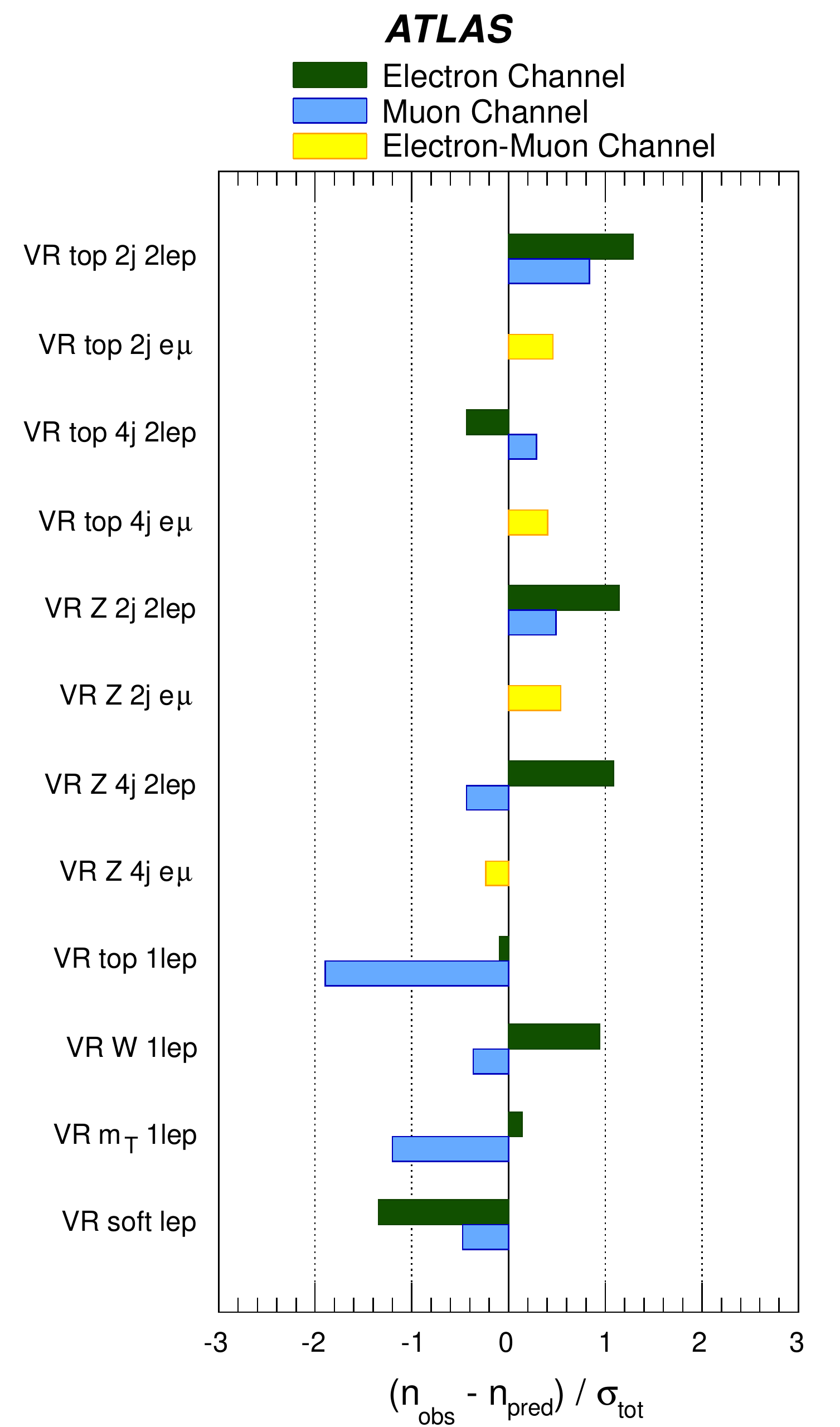}
\caption{Example produced by the ATLAS collaboration and taken from Ref.~\cite{PhysRevD.86.092002}. Summary of the fit results in the validation regions. 
The difference between the observed and predicted number, divided by the total (statistical and systematic) uncertainty on the prediction, is shown for each validation region.}
\label{fig:pull_1L}
\end{figure}

Other validation plots can be produced, mostly helpful for internal or debugging purposes.
Likelihood scans can be made for any of the fit parameters to help understand the likelihood maximization
performed in the fit.
Furthermore, the correlation matrix of any fit can be plotted 
to study correlations between the fit parameters and possible degenerate degrees of freedom. 
Those examples are not unique to HistFitter, and are therefore not shown here.

\subsection{Scripts for event yield, systematic uncertainty and pull tables}

The production of detailed tables showing the estimated background event levels, the number of observed events and the breakdown 
of the systematic uncertainties is an essential part of every analysis. 
HistFitter includes several scripts to produce publication ready (LaTeX) tables.

Table~\ref{tab:yields_0L} shows the results of the background-only fit to the CRs, extrapolated to a set of SRs
and broken down into the various background processes, as taken from Ref.~\cite{PhysRevD.87.012008} and produced with HistFitter. 
The total background prediction, combined with the number of observed events in a signal region, 
allows 
the discovery $p$-value or limit setting to be re-derived by others, externally from the original analysis.
The background predictions before the fit are shown in parenthesis.
The error on the total background estimate shows the statistical (from limited MC simulation and CR statistics combined) 
and systematic uncertainties separately, 
while for the individual background samples the combined uncertainties are given as a single number. 
The uncertainties on the predicted background event yields are quoted as symmetric, 
except where the negative error reaches below zero predicted events, in which case the negative error gets truncated to zero.
The errors shown are the after-fit uncertainties, though before-fit uncertainties can also be shown by the table-production script.

\begin{table}[h!]
\begin{center}
{\footnotesize
\begin{tabular}{|c|c|c|c|c|c|}
\hline
\multirow{2}{*}{Process}     &\multicolumn{5}{|c|}{Signal Region} \\ 
\cline{2-6}
& SR-A tight & SR-B tight & SR-C tight & SR-D tight & SR-E tight \\
\hline
\hline
$t\bar{t}$+single top  &  $0.2\pm 0.2$ (0.1) &  $0.3\pm 0.3$ (0.2) &  $2.0\pm 1.5$ (1.2) &  $2.4\pm 1.7$ (1.4) &  $4.2\pm 4.7$ (3.0) \\
$Z$+jets  &  $3.3\pm 1.5$ (4.0) &  $2.0\pm 1.3$ (2.1) &  $2.0\pm 1.0$ (5.6) &  $0.9\pm 0.6$ (3.4) &  $3.4\pm 1.6$ (2.3) \\
$W$+jets  &  $2.2\pm 1.0$ (1.9) &  $1.0\pm 0.6$ (0.8) &  $1.5\pm 1.3$ (2.7) &  $2.4\pm 1.4$ (2.5) &  $2.8\pm 1.9$ (1.5) \\
Multi-jets   &  $0.00\pm 0.02$ (0.01) &  $0.00\pm 0.07$ (0.02) &  $0.00\pm 0.03$ (0.01) &  $0.0\pm 0.3$ (0.1) &  $0.5\pm 0.4$ (0.9) \\
Di-bosons &  $1.8\pm 0.9$ (2.0) &  $1.8\pm 0.9$ (1.9) &  $0.5\pm 0.3$ (0.5) &  $2.2\pm 1.1$ (2.2) &  $2.5\pm 1.3$ (2.5) \\
Total  & $7.4\pm 1.3\pm 1.9$ & $5.0\pm 0.9\pm 1.7$ & $6.0\pm 1.0\pm 2.0$ & $7.8\pm 1.0\pm 2.4$ & $13\pm 2\pm 6$ \\
Data  & 1 & 1 & 14 & 9 & 13 \\
\hline
\end{tabular}
}
\caption{\label{tab:yields_0L} 
Example produced by the ATLAS collaboration and taken from Ref.~\cite{PhysRevD.87.012008}. 
Illustration of observed numbers of events in data and fitted background components in each SR, as obtained 
from a background-only fit to CRs. For the total
background estimates, the quoted uncertainties give the statistical (MC simulation and CR combined) and systematic
uncertainties respectively. For the individual background components, the total
uncertainties are given, while the values in parenthesis indicate the pre-fit predictions. 
}
\end{center}
\end{table}

There are two methods implemented in HistFitter to calculate the systematic uncertainty 
on a background level prediction of an analysis
associated to a specific (set of) nuisance parameter(s), 
such as detector response effects or theoretical uncertainties. 
\begin{enumerate}
\item The first method takes the nominal after-fit result and sets all floating parameters constant. 
Then, iteratively, it sets each (or several, as requested) nuisance parameter(s) $\eta_i$ floating, 
and calculates the uncertainty propagated to the background prediction due to the specific parameter(s), 
using the covariance matrix of the nominal fit and Eq.~\ref{eq:errorProp}.
\item The second method sets a single (or multiple, as requested) floating nuisance parameter(s) constant 
and then refits the data, thus excluding these systematic uncertainties from the model. 
The quadratic difference between the total error of the nominal setup and the fixed parameter(s) setup 
is then assigned as the systematic uncertainty, as follows:
\begin{equation}
\sigma_{\eta_1} = \sqrt{\Big(\sigma_\mathrm{tot}^\mathrm{nominal}\Big)^2 - \Big(\sigma_\mathrm{tot}^{\eta_1=C}\Big)^2 }\,.
\end{equation}
\end{enumerate}
Table~\ref{table:syst_2LSS} shows the systematic breakdown of the background estimate uncertainty 
in a set of signal regions, as produced with method one and taken from Ref.~\cite{Aad:2014pda}. 
Each row shows the uncertainty corresponding to one or more nuisance parameters, as detailed in the reference.

\begin{table}[h!]
\begin{center}
\setlength{\tabcolsep}{0.0pc}
{\small
\begin{tabular*}{\textwidth}{@{\extracolsep{\fill}}lrrrrr}
\noalign{\smallskip}\hline\noalign{\smallskip}
{\bf Signal Region}           & SR3b & SR0b            & SR1b            & SR3Llow            & SR3Lhigh             \\[-0.05cm]
\noalign{\smallskip}\hline\hline\noalign{\smallskip}
{\bf Observed events}           & $1$  & $14$              & $10$                         & $6$              & $2$                    \\
\noalign{\smallskip}\hline\noalign{\smallskip}
{\bf Total expected background events}  & $2.2 \pm 0.8$ & $6.5 \pm 2.3$          & $4.7 \pm 2.1$                  & $4.3 \pm 2.1$          & $2.5 \pm 0.9$              \\
\noalign{\smallskip}\hline\hline\noalign{\smallskip}
\multicolumn{5}{l}{\bf Systematic uncertainties on expected background} \\
\noalign{\smallskip}\hline\noalign{\smallskip}
Fake-lepton background         & $\pm 0.6$  & $_{-1.2}^{+1.5}$          &$_{-0.8}^{+1.2}$           & $\pm 1.6$          & $<0.1$      \\
Theory unc. on dibosons       & $<0.1$     & $\pm 1.5$          & $\pm 0.3$                & $\pm 0.4$          & $\pm 0.4$       \\
Jet and \met\ scale and resolution         & $\pm 0.1$  & $\pm 0.7$          & $\pm 0.4$                  & $\pm 0.4$          & $\pm 0.3$       \\
Monte Carlo statistics        & $\pm 0.1$     & $\pm 0.5$          & $\pm 0.2$               & $\pm 0.4$          & $\pm 0.4$       \\
$b$-jet tagging          & $\pm 0.2$  & $\pm 0.5$          & $\pm 0.1$                & $<0.1$          & $\pm 0.1$       \\
Theory unc. on  $\ttbar V$, $\ttbar H$, $tZ$ and $t\bar{t}t\bar{t}$         & $\pm 0.4$  & $\pm 0.3$     & $\pm 1.7$         & $\pm 1.0$        & $\pm 0.6$       \\
Trigger, luminosity and pile-up         & $<0.1$      & $\pm 0.1$          & $\pm 0.1$             & $\pm 0.1$          & $\pm 0.1$       \\
Charge-flip background           & $\pm 0.1$       & $\pm 0.1$          & $\pm 0.1$          &  --          & --       \\
Lepton identification            & $<0.1$    & $\pm 0.1$          & $<0.1$            & $\pm 0.1$          & $\pm 0.1$       \\
\noalign{\smallskip}\hline\hline\noalign{\smallskip}
\end{tabular*}
}\end{center}
\caption{ Example produced by the ATLAS collaboration and taken from Ref.~\cite{Aad:2014pda}. Number of observed data events and expected backgrounds and
  summary of the systematic uncertainties on the background predictions for
 SR3b, SR0b, SR1b, SR3Llow and SR3Lhigh.
  The breakdown of the systematic uncertainties on the expected
  backgrounds, expressed in units of events, is also shown.
  The individual uncertainties are correlated and therefore do not
  necessarily add up in quadrature to
  the total systematic uncertainty.}
\label{table:syst_2LSS}
\end{table}

\section{Interpretation of results}
\label{sec:interpret}

HistFitter provides the functionality to perform hypothesis tests of the data 
through calls to the appropriate RooStats classes, and to interpret the corresponding results in the form of plots and tables. 
Four different statistical tests are available in HistFitter.
Each of these depend on the fit setups outlined in Sec.~\ref{sec:fit:bkg-dis-exc}.
In each of these setups both the CR(s) and SR(s) are part of the input to the fit.

In the absence of an observed excess of events in one or more SR(s), 
the first two methods set exclusion limits on specific signal models.
Both use the model-dependent signal fit configuration.
The third approach obtains exclusion upper limits on any potential new physics signal, without model dependency. 
The fourth interpretation performs the significance determination of a potentially observed event excess.
Both of these rely on the model-independent signal fit configuration. 

These different statistical tests are discussed in the following sections.
As discussed in Sec.~\ref{sec:roostats}, a Frequentist approach is used in all of the methods explained below,
together with the CLs method in case of exclusion hypothesis tests.

In Sec.~\ref{sec:fit:bkg-dis-exc} we have introduced background-level estimates that are 
obtained from a background-only fit performed in the CRs only.
For completeness we also introduce the concept of background-level estimates
obtained from a background-only fit in both the CRs and 
SRs\footnote{Sometimes confusingly called ``unblinded'' background-level estimates.}.
These fit results are obtained with any model-(in)dependent signal fit setup, 
where the signal component has been turned off.
As discussed in Sec.~\ref{sec:roostats}, the RooStats routines employed here, using the profile 
likelihood ratio as test statistic,
perform such a background-only fit to both the CRs and SRs before running the $p$-value determination;
a strategy using the most accurate background-level estimates available.
A consequence of this is discussed in Sec.~\ref{sec:modindepul}, together with an example.

All tables and figures shown in this section (except for Tab.~\ref{tab:ULexpected}) 
come directly from publications by the ATLAS collaboration 
and serve only as illustrations of the HistFitter tools that are discussed.

\subsection{Signal model hypothesis test}
\label{signalhypotest}

The various tools and scripts to execute the signal hypothesis tests,
as well as to visualize the results, provided in  HistFitter are explained here. 

In the signal model hypothesis test,
a specific model of new physics is tested against the background-only model assumption. 
A signal model prediction is present in all CRs and SRs, 
as implemented in the model-dependent signal limit fit configuration of Sec.~\ref{sec:fit:bkg-dis-exc}. 
The parameter of interest used in these hypothesis tests is the signal strength parameter, 
where a signal strength of zero corresponds to the background-only model, 
and a signal strength of one to the background plus signal model. 

A fit of the background plus signal model is performed first, 
with the signal strength being a free normalization parameter, 
to obtain an idea about potential fit failures or problems in the later hypothesis testing. 
The fit result is stored for later usage in the interpretation of the hypothesis test results.

Usually, signal hypothesis tests are run for multiple signal scenarios making up a specific model grid, 
e.g. by modifying a few parameters for a specific supersymmetry model. 
HistFitter provides the possibility to collect the results for the different signal scenarios in a data text file, 
collecting in particular the observed and expected CLs values, but also the $p$-values for the various signals. 
Only results of hypothesis tests with a successful initial free fit are saved to the data text file. 
Another macro transforms these entries into two-dimensional histograms, 
for example showing the CLs values versus the SUSY parameter values (or particle masses) of the signal scenarios tested.
A linear algorithm is used to interpolate the CLs values between signal model parameter values.

HistFitter provides macros to visualize the results of the hypothesis tests graphically. 
An example is shown in Fig.~\ref{fig:interpret:contour} (left), taken from Ref.~\cite{PhysRevD.86.092002}.
The exclusion limits are shown at 95\% confidence level, based on the CLs prescription,
in a so-called one decay step (1-step) simplified model~\cite{Alves:2011wf}.
There are only two free parameters in these particular SUSY models, $m_{\tilde{g}}$ and $m_{\tilde{\chi}_1^0}$,
which are used as the variables on the axes to represent this specific SUSY model grid. 
The dark dashed line indicates the expected limit as function of gluino and neutralino masses 
and the solid red line the observed limit. 
The yellow band gives the $1 \sigma$ uncertainty on the expected limit, excluding the 
theoretical uncertainties on the signal prediction. 
The dotted red lines show the impact of the theoretical uncertainties of the signal model prediction 
on the observed exclusion contour.

Likewise Fig.~\ref{fig:interpret:contour} (right) shows the observed and expected exclusion limits on a gluino-mediated
 top squark production model, taken from Ref.~\cite{Aad:2014pda},
as obtained from the statistical combination of four multi-binned SRs performed with HistFitter.
Besides the expected exclusion limit from the simultaneous fit to all SRs, 
the expected exclusion limits from the individual SRs are shown for comparison.

\begin{figure}
 \centering
 \includegraphics[width=0.48\textwidth]{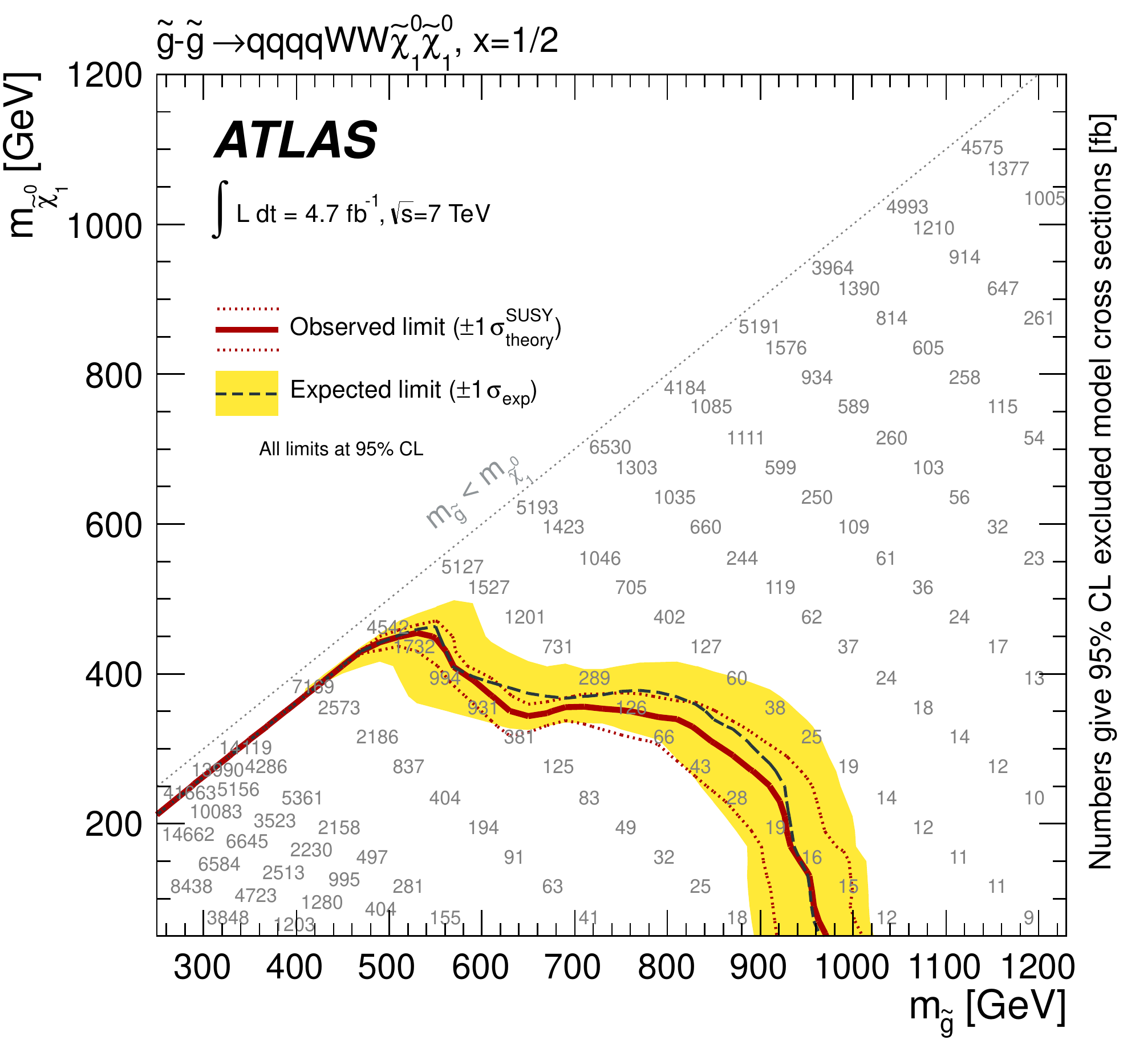}
 {\hspace{0.15cm}}
 \includegraphics[width=0.48\textwidth]{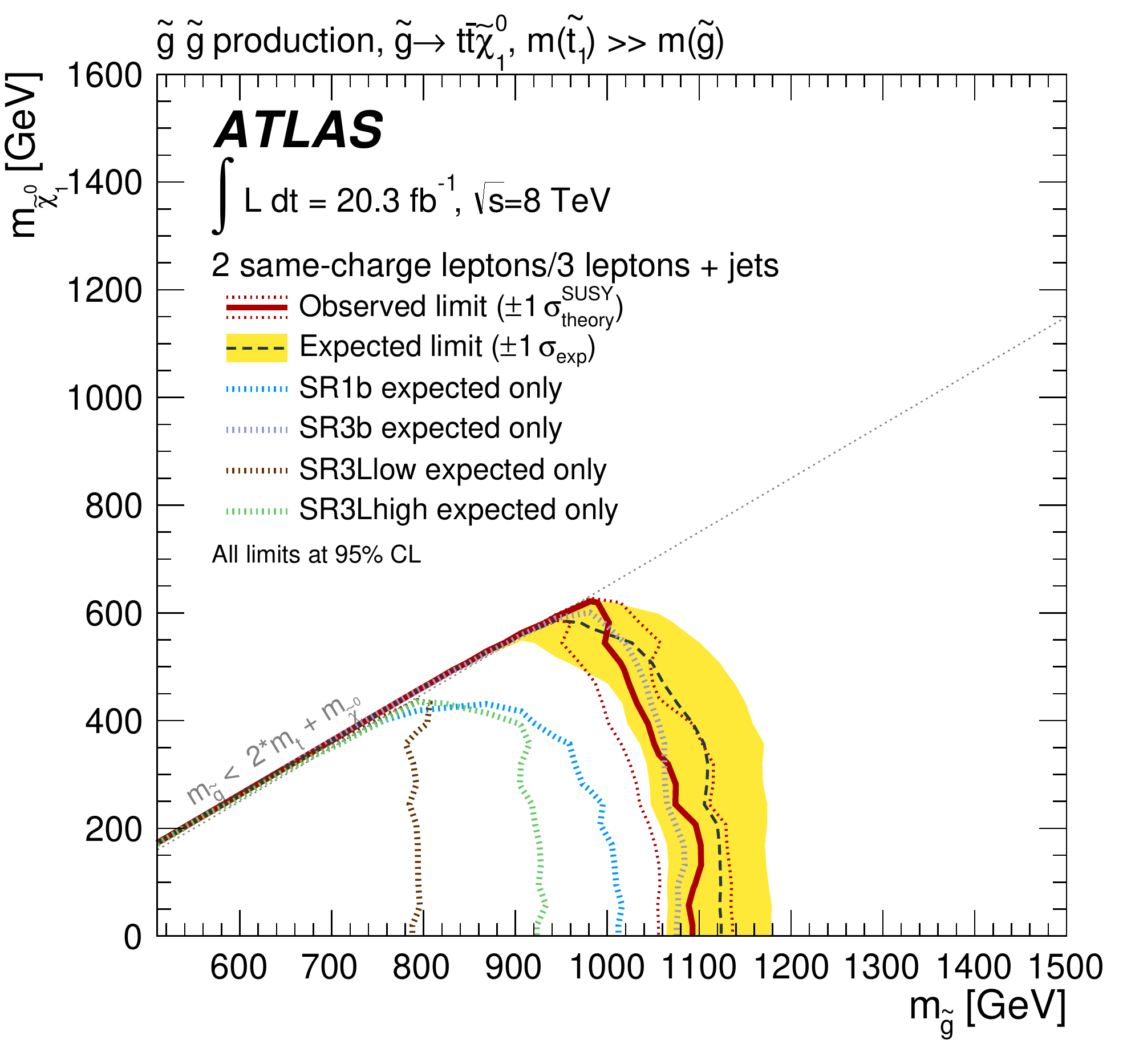}
 \caption{Examples produced by the ATLAS collaboration. Excluded regions at 95\% confidence level in a 1-step simplified model (left), with initial gluino pair production and subsequent decay of the gluinos via $\tilde{g} \rightarrow qq\tilde{\chi}^{\pm}_1 \rightarrow qqW\tilde{\chi}_1^0$ and taken from Ref.~\cite{PhysRevD.86.092002}.
Observed and expected limits on gluino-mediated top squark production (right), obtained from a simultaneous fit to four signal regions, and expected exclusion limits from the individual signal regions, produced by the ATLAS collaboration and taken from Ref.~\cite{Aad:2014pda}.}
\label{fig:interpret:contour}
\end{figure}

\subsection{Signal strength upper limit}
\label{signalupperlimit}

As in Sec.~\ref{signalhypotest}, we consider a specific signal model and the model-dependent signal limit fit 
configuration in this section.
HistFitter provides the possibility to set an upper limit on the signal strength parameter $\mu_{\rm sig}$
given the observed data in the signal regions. 
To do so, the value of the signal strength needs to be evaluated for 
which the CLs value falls below a certain level, usually 5\% (for a 95$\%$ CL upper limit).

In an initial scan, multiple hypothesis tests are executed using the asymptotic calculator~\cite{Cowan:2010js} 
to evaluate the CLs values for a wide range of signal strength values. 
A second scan follows in a smaller, refined interval, using the expected upper limit derived from the first scan. 

The obtained upper limit on the signal strength can then easily be converted into 
an upper limit on the excluded cross section of the signal model tested initially. 
These cross section upper limits are often displayed together with the limits obtained from the signal hypothesis test, 
introduced in Sec.~\ref{signalhypotest}. 
The example discussed in Fig.~\ref{fig:interpret:contour} (left) includes 
the cross section upper limits as grey numbers for each of the tested signal models.

\subsection{Model-independent upper limit}
\label{sec:modindepul}

To obtain the 95\% CL upper limit on the number of events in a ``beyond the Standard Model prediction'' 
for each SR,
the fit in the SR proceeds in the same way as the background-only fit, 
except that the number of events observed in the signal region (evaluated in one bin) 
is added as an input to the fit.
The signal strength parameter is constrained to be non-negative.
The statistical test uses the model-independent signal fit configuration described in Sec.~\ref{sec:fit:bkg-dis-exc}.
This model-independent upper limit 
is evaluated using the same approach as in Sec.~\ref{signalupperlimit}. 

By normalizing the signal-strength from the fit to the integrated luminosity of the data sample, and accounting for the uncertainty on the recorded luminosity, this
can be interpreted as the upper limit on the visible cross section of new physics, $\sigma_{\rm vis}$.
Here $\sigma_{\rm vis}$ is defined as the product of acceptance, reconstruction efficiency and production cross section.

HistFitter includes a script to calculate and present the upper limits on the number of signal events 
and on the visible cross section 
in a publication ready (LaTeX) table. 
An example, based on the background estimates of Tab.~\ref{table:syst_2LSS}, 
is shown in Tab.~\ref{tab:interpret:modelind}.

\begin{table}
\begin{center}\setlength{\tabcolsep}{0.0pc}
\begin{tabular*}{\textwidth}{@{\extracolsep{\fill}}lcccc}
\noalign{\smallskip}\hline\noalign{\smallskip}
{\bf Signal channel}                        & $\langle \sigma_{\rm vis}\rangle_{\rm obs}^{95}$[fb]  &  $S_{\rm obs}^{95}$  & $S_{\rm exp}^{95}$ & $p(s=0)$\\
\noalign{\smallskip}\hline\noalign{\smallskip}
SR3b    &  $0.19$ &  $3.9$  & ${4.4}^{+1.7}_{-0.6}$ & $0.50$\\
SR0b    &  $0.80$ &  $16.3$ & ${8.9}^{+3.6}_{-2.0}$ & $0.03$\\
SR1b    &  $0.65$ &  $13.3$ & ${8.0}^{+3.3}_{-2.0}$ & $0.07$\\
SR3Llow &  $0.42$ &  $8.6$  & ${7.2}^{+2.9}_{-1.3}$ & $0.29$\\
SR3Lhigh&  $0.23$ &  $4.6$  & ${5.0}^{+1.6}_{-1.1}$ & $0.50$\\
\noalign{\smallskip}\hline\noalign{\smallskip}
\end{tabular*}
\end{center}
\caption{ Example produced by the ATLAS collaboration and taken from Ref.~\cite{Aad:2014pda}.
The 95$\%$ CL upper limits on the visible cross section ($\langle \sigma_{\rm
vis}\rangle_{\rm obs}^{95}$), defined as the product of acceptance,
reconstruction efficiency and production cross section, and the observed and
expected 95$\%$ CL upper limits on the number of signal events ($S_{\rm
obs}^{95}$ and $S_{\rm exp}^{95}$). 
The last column shows the probability, capped at $0.5$, that a background-only experiment 
is more signal-like than observed number number of events in a signal region (discussed in Sec.~\ref{sec:discpvalue}).  
}
\label{tab:interpret:modelind}
\end{table}

As discussed in Sec.~\ref{sec:roostats}, the profile-likelihood based 
hypothesis tests use the background-level estimates 
obtained from a background-only fit to both the CRs and SRs 
(the best estimates available).
For consistency, both the observed and expected upper limit (or $p$-value) determination use 
the same background-level estimates, 
such that the expected limit is the most compatible and predictive assessment 
for the observed limit. 
As a consequence, the expected upper limit depends indirectly on the observed data.

This feature is demonstrated in Tab.~\ref{tab:ULexpected}, which shows a counting experiment
with a constant background expectation and an increasing number of observed events,
resulting in a consistent rise in the internal background-level estimates.
As a result, the 95\% CL upper limit on the expected number of signal events 
rises as a function of the number of observed events.
This behavior, though perhaps counter-intuitive, 
is a consequence of the profile-likelihood based limit setting procedure employed here.

\begin{table}[!th]
\begin{center}\setlength{\tabcolsep}{0.0pc}
\begin{tabular*}{\textwidth}{@{\extracolsep{\fill}}ccccc}
\noalign{\smallskip}\hline\noalign{\smallskip}
Expected background  & Observed events  &  Background estimate  &  $S_{\rm obs}^{95}$  & $S_{\rm exp}^{95}$   \\
\noalign{\smallskip}\hline\noalign{\smallskip}
$3.7 \pm 1.6$  &  $0$   &  $1.9 \pm 0.9$ &  $2.6$   & ${4.4}^{+2.5}_{-1.5}$   \\
$3.7 \pm 1.6$  &  $4$   &  $3.8 \pm 1.2$ &  $6.3$   & ${6.1}^{+2.9}_{-1.9}$  \\
$3.7 \pm 1.6$  &  $8$   &  $5.3 \pm 1.5$ &  $10.8$  & ${6.6}^{+3.3}_{-2.0}$  \\
$3.7 \pm 1.6$  &  $15$  &  $8.7 \pm 2.2$ &  $19.2$  & ${8.7}^{+3.8}_{-2.5}$  \\
\noalign{\smallskip}\hline\noalign{\smallskip}
\end{tabular*}
\end{center}
\caption{
The observed and expected 95$\%$ CL upper limits on the number of signal events ($S_{\rm obs}^{95}$ and $S_{\rm exp}^{95}$),
as a function the background expectation and the observed number of events, 
as obtained with asymptotic formulas for a single-bin counting experiment.
The third column shows the background estimate 
obtained from a fit to the expected background and observed number of events.
\label{tab:ULexpected}}
\end{table}

\subsection{Background-only hypothesis test}
\label{sec:discpvalue}

For completeness, yet not tailored to HistFitter needs,
the background-only hypothesis test quantifies the significance of an excess of events in the 
signal region by the probability that a background-only experiment is more signal-like than observed, also called the discovery $p$-value.
The same fit configuration is used as in Sec.~\ref{sec:modindepul}.
An example of calculated discovery $p$-values is shown in the last column of Tab.~\ref{tab:interpret:modelind}.
The probability of the SM background to fluctuate to the observed number of events or higher in each SR
has been capped at $0.5$.

\section{Public release}
\label{sec:implementation}

The HistFitter software package is publicly available\footnote{Support is provided on a best-effort basis.} through the web-page \url{http://cern.ch/histfitter},
which requires \texttt{ROOT} release v5.34.20 or greater. 
The web-page contains a description of the source code, a tutorial on how to set up an analysis, 
and working examples of how to run and use the code.

\section{Conclusion}
\label{sec:conclu}

We have presented a software framework for statistical data analysis, called HistFitter, 
that has been used extensively by the ATLAS Collaboration to analyze big datasets 
originating from proton-proton collisions at the LHC at CERN. 

HistFitter provides a programmable framework to build and test a set of data models of nearly arbitrary complexity.
Starting from an input configuration, defined by users, 
it uses the software packages HistFactory, RooStats, RooFit and ROOT
to construct PDFs that are fitted to data
and interpreted with statistical tests, automatically.

HistFitter brings forth several innovative features. 
It provides a modular configuration interface with a trickle-down mechanism 
that is very efficient and intuitive for users.
It has built-in concepts of control, signal and validation regions, 
with rigorous statistical treatment, tailored to support a complete 
particle physics analysis.
It is capable of working with multiple data models at once,
which introduces an additional level of abstraction
that is powerful when searching for new phenomena in large experimental datasets.
Finally, HistFitter provides a sizable collection of tools and options,
resulting from experience gained during real-life analysis of ATLAS Run-1 data,
that allows, through simple command-line commands,
the presentation of end results with publication-style quality.

\section*{Acknowledgments}
\label{sec:Acknowledgments}

We are grateful to the RooFit, RooStats, and HistFactory authors for a fruitful collaboration and useful feedback,
in particular to Kyle Cranmer, Lorenzo Moneta and Wouter Verkerke.
We would like to thank our ATLAS colleagues for permission to reproduce the published figures and tables 
in Secs.~\ref{sec:pres} and~\ref{sec:interpret} to illustrate the HistFitter tools discussed.
We also thank the ATLAS collaboration and its SUSY physics group
for useful discussions and suggestions for the development of HistFitter,
and in particular thank Monica D'Onofrio and Jamie Boyd for their feedback on this paper.
And we are specifically grateful to following members of the ATLAS SUSY
physics group for their support and contributions to the development of
HistFitter: Andreas Hoecker, Till Eifert, Zachary Marshall, Emma Sian Kuwertz, Evgeny Khramov, Sophio Pataraia and Marcello
Barisonzi.

This work was supported by CERN, Switzerland; 
the DFG cluster of excellence ``Origin and Structure of the Universe''; 
the Natural Sciences and Engineering Research Council of Canada and the ATLAS-Canada Subatomic Physics Project Grant; 
the Department Of Energy and the National Science Foundation of the United States of America; 
FOM and NWO, the Netherlands; and STFC, United Kingdom.

\clearpage
\appendix
\begin{appendix}

\section{Example configuration}
\label{app:exampleconfiguration}

An example HistFitter configuration file is shown here,
using the programmable components of Sec.~\ref{sec:pdf}. A description of the example follows below.
The example configuration illustrates a single-bin counting experiment. In short:
\begin{itemize}
\item The single defined channel is called \texttt{SR}.
\item This contains two background samples \texttt{A} and \texttt{B}, besides the data sample \texttt{Data}. 
All required inputs are extracted from the files \texttt{fileA.root} and \texttt{fileB.root}.
\item There are two systematic uncertainties defined, \texttt{treeSys} and \texttt{weightSys}, where
the latter is only applied to sample \texttt{A}.
There is also a luminosity uncertainty, applied to both samples.
\item There are two fit configuration objects created, 
one for a discovery hypothesis test,  labeled \texttt{Discovery},
and one for a model-dependent exclusion fit, labeled \texttt{Exclusion}.
These contain a (dummy) signal sample (predicting 1 signal event) 
and a specific signal sample, called \texttt{Signal}, respectively.
\end{itemize}
Additional options and comments are given in-line. \\

\lstset{basicstyle=\footnotesize\ttfamily,breaklines=true}
\begin{lstlisting}[language=Python]
from configManager import configMgr
from ROOT import kBlack, kGreen, kAzure, kMagenta, kPink
from configWriter import fitConfig, Measurement, Channel, Sample
from systematic import Systematic

# Parameters for hypothesis test and upper limit
configMgr.calculatorType = 2 # Asymptotic
configMgr.testStatType = 3 # Frequentist
configMgr.nPoints = 20 # #points to use in UL scan

# Now we start to build the data model
# First define HistFactory attributes
configMgr.analysisName = "MyOneBinExample"
configMgr.histCacheFile = "data/"+configMgr.analysisName+".root"
configMgr.outputFileName = "results/"+configMgr.analysisName+"_Output.root"

# Scaling calculated by outputLumi / inputLumi
configMgr.inputLumi = 0.001 # Luminosity of input after weighting
configMgr.outputLumi = 4.713 # Luminosity required for output histograms
configMgr.setLumiUnits("fb-1")
configMgr.blindSR = False
configMgr.useSignalInBlindedData = False 

# Set the files to read from, use the intermediate cache file if not reading trees
bgdFiles = []
if configMgr.readFromTree:
    bgdFiles.append("fileA.root")
    bgdFiles.append("fileB.root")
else:
    bgdFiles = [configMgr.histCacheFile]
    pass
configMgr.setFileList(bgdFiles)

# Dictionary of cuts to place on input trees
configMgr.cutsDict["SR"] = "leptonPt > 20 && ( (met > 160 && met/meff > 0.2) || met > 1000)"

# Tuples of nominal weights 
configMgr.weights = ("genWeight", "eventWeight", "leptonWeight", "triggerWeight")

# List of systematics
# Weight-based: provide up and down weights, and a systematic type
highWeights = ("eventWeight", "weightUp")
lowWeights = ("eventWeight", "weightDown")
weightSys = Systematic("KtScaleTop", configMgr.weights, highWeights, lowWeights, "weight", "overallSys")

# Tree-based: provide name, plus 3 suffixes for nominal, up and down, and a systematic type 
treeSys = Systematic("SYS", "_NoSys", "_SYS_up", "_SYS_down", "tree", "overallSys")
configMgr.nomName = "_NoSys"

# List of samples and their plotting colours
aSample = Sample("A", kGreen)
bSample = Sample("B", kAzure)
dataSample = Sample("Data", kBlack)
dataSample.setData()

# Signal model independent fit config (aka Discovery fit)
# -------------------------------------------------------
# Fit config instance
discoveryFitConfig = configMgr.addFitConfig("Discovery")
meas = discoveryFitConfig.addMeasurement(name="NormalMeasurement", lumi=1.0, lumiErr=0.039)
meas.addPOI("mu_SIG")

# Samples
discoveryFitConfig.addSamples([aSample, bSample, dataSample])

# Systematics: add weightSys to one sample, treeSys to all
discoveryFitConfig.getSample("A").addSystematic(weightSys)
discoveryFitConfig.addSystematic(treeSys)

# Channel with a dummy sample of 1
SR = discoveryFitConfig.addChannel("cuts", ["SR"], 1, 0.5, 1.5)
discoveryFitConfig.setSignalChannels([SR])
SR.addDiscoverySamples(["SIG"], [1.], [0.], [100.], [kMagenta])

# Signal model dependent fit config (aka Exclusion fit)
# ----------------------------------------------------
# Fit config instance
exclusionFitConfig = configMgr.addFitConfig("Exclusion")
meas = exclusionFitConfig.addMeasurement(name="NormalMeasurement", lumi=1.0, lumiErr=0.039)
meas.addPOI("mu_SIG")

exclusionFitConfig.addSamples([aSample, bSample, dataSample])

exclusionFitConfig.getSample("A").addSystematic(weightSys)
exclusionFitConfig.addSystematic(treeSys)

SR = exclusionFitConfig.addChannel("cuts", ["SR"], 1, 0.5, 1.5)
exclusionFitConfig.setSignalChannels([SR])

# Add a real signal sample to test
sigSample = Sample("Signal", kPink)
sigSample.setFileList(["signal.root"])
sigSample.setNormByTheory()
sigSample.setNormFactor("mu_SIG", 1., 0., 5.)                    
exclusionFitConfig.addSamples(sigSample)
exclusionFitConfig.setSignalSample(sigSample)

\end{lstlisting}
\label{list:examplecode}

\end{appendix}

\clearpage
\addcontentsline{toc}{section}{References}
\bibliography{References}

\end{document}